\documentclass[letterpaper,twocolumn,10pt]{article}
\usepackage{usenix2019_v3}

\usepackage[compact]{titlesec}
\usepackage[T1]{fontenc}   %
\usepackage{tcolorbox}

\usepackage{graphicx}
\usepackage{multirow}
\usepackage{amsmath}
\usepackage[font=bf]{caption}

\usepackage{subcaption}
\usepackage{minted}
\usepackage{listings}
\usepackage{xcolor}
\usepackage{xurl}
\usepackage{booktabs}

\usepackage[ruled,vlined]{algorithm2e}

\definecolor{functioncolor}{RGB}{170, 51, 119}  %
\definecolor{arrowcolor}{RGB}{255, 69, 0}       %

\begin{document}
\date{}
\title{\sysname:  Low Cost   and High Performance  Metrics Observability at Scale}

\author{
Milind Srivastava\textsuperscript{1},
Zeying Zhu\textsuperscript{2},
Peilin Xin\textsuperscript{1},
Yancheng Yuan\textsuperscript{2},
Zaoxing Liu\textsuperscript{2},
Vyas Sekar\textsuperscript{1}
\\
\textsuperscript{1}Carnegie Mellon University
\quad
\textsuperscript{2}University of Maryland \\ 
}

\newcommand\Mo{Metrics observability\xspace}
\newcommand\mo{MO\xspace}

\newcommand{\molong}{metrics observability\xspace}

\newcommand\olly{observability\xspace}
\newcommand\Olly{Observability\xspace}

\newcommand\Prom{Prometheus\xspace}
\newcommand\PromQL{PromQL\xspace}
\newcommand\VM{VictoriaMetrics\xspace}
\newcommand\Signoz{Signoz\xspace}
\newcommand\Grafana{Grafana\xspace}
\newcommand\Clickhouse{ClickHouse\xspace}

\newcommand\Deathstarbench{Deathstarbench\xspace}

\newcommand\container{container\xspace}
\newcommand\containers{containers\xspace}

\newcommand\uc{usecase\xspace}
\newcommand\ucs{usecases\xspace}

\newcommand\metric{metric\xspace}
\newcommand\metrics{metrics\xspace}
\newcommand\Metric{Metric\xspace}
\newcommand\Metrics{Metrics\xspace}

\newcommand\query{query\xspace}
\newcommand\queries{queries\xspace}

\newcommand\dataMath{\ensuremath{D}}
\newcommand\dataVectorMath{\ensuremath{\vec{D}}}

\newcommand\qasMath{\ensuremath{QAS}}
\newcommand\qasVectorMath{\ensuremath{\vec{QAS}}}

\newcommand\sdsMath{\ensuremath{SDS}}
\newcommand\sdsVectorMath{\ensuremath{\vec{SDS}}}

\newcommand\queryMath{\ensuremath{Q}}
\newcommand\queryVectorMath{\ensuremath{\vec{Q}}}
\newcommand\queryVectorEnsembleMath{\ensuremath{\{\vec{Q}\}}}

\newcommand\statusQuoFuncMath{\ensuremath{f_Q}}

\newcommand\summaryChainMath{\ensuremath{s}}
\newcommand\summarizeFuncMath{\ensuremath{S}}
\newcommand\CandidateSummaryChainSetMath{\ensuremath{CS}}

\newcommand\estimateDAGMath{\ensuremath{e}}
\newcommand\estimateFuncMath{\ensuremath{E}}

\newcommand\ingestCostMath{\ensuremath{ingest\_cost}}
\newcommand\queryCostMath{\ensuremath{query\_cost}}

\newcommand\PlanMath{\ensuremath{p}}
\newcommand\PlanSetMath{\ensuremath{P}}

\newcommand\decisionDeploySummaryChain{\ensuremath{x}\xspace}
\newcommand\decisionMapPairToAQE{\ensuremath{y}\xspace}

\newcommand\accuracyFuncMath{\ensuremath{acc}}
\newcommand\latencyFuncMath{\ensuremath{lat}}
\newcommand\feasibilityFuncMath{\ensuremath{feasible}}

\newcommand\frequencyRQEMath{\ensuremath{f}}

\newcommand{\Timevar}{\ensuremath{\mathit{T}}}
\newcommand{\repeating}{\ensuremath{\mathit{\delta}}}
\newcommand{\lookback}{\ensuremath{\Delta}}

\newcommand\RQEMath{\ensuremath{r}}
\newcommand\RQESetMath{\ensuremath{R}}

\newcommand\AQEMath{\ensuremath{a}}
\newcommand\AQESetMath{\ensuremath{A}}

\newcommand\accuracySLAMath{\ensuremath{\epsilon}}
\newcommand\latencySLAMath{\ensuremath{L}}

\newcommand{\insertCpu}{\mathsf{insertCPU}\xspace}
\newcommand{\mergeCpu}{\mathsf{mergeCPU}\xspace}
\newcommand{\subtractCpu}{\mathsf{subtractCPU}\xspace}
\newcommand{\estimateCpu}{\mathsf{estimateCPU}\xspace}

\newcommand\SummaryOpInstance{\ensuremath{algo}\xspace}

\definecolor{ForestGreen}{RGB}{34,139,34}

\newcommand{\deftt}[2]{\expandafter\newcommand\csname #1\endcsname{\texttt{#2}\xspace}}
\newcommand{\defttself}[1]{\deftt{#1}{#1}}

\newcommand\sysname{TurboProm\xspace}

\newcommand\summary{synopsis\xspace}
\newcommand\summaries{synopses\xspace}
\newcommand\Summary{Synopsis\xspace}
\newcommand\Summaries{Synopses\xspace}

\newcommand\Summarizer{Summarizer\xspace}
\newcommand\QueryEngine{QueryEngine\xspace}
\newcommand\StorageEngine{StorageEngine\xspace}
\newcommand\Planner{Planner\xspace}

\deftt{M}{Metric}
\deftt{Ms}{Metrics}

\defttself{Subpopulation}
\defttself{Subpopulations}

\newcommand{\SummaryOp}{{\tt {\Summary}Algo}\xspace}
\newcommand{\SummaryLib}{{\tt {\Summary}Lib}\xspace}

\newcommand{\SummaryChain}{{\tt {\Summary}Plan}\xspace}
\newcommand{\SummaryChains}{{\tt {\Summary}Plans}\xspace}
\newcommand\EstimationDAG{{\tt EstimationPlan}\xspace}
\newcommand\EstimationDAGs{{\tt EstimationPlans}\xspace}
\newcommand\basicEstimationDAG{{\tt BasicEstimationDAG}\xspace}

\newcommand\sdsFull{\texttt{self-describing \summary}\xspace}
\newcommand\SDSFull{\texttt{Self-Describing \Summary}\xspace}
\newcommand\sdssFull{\texttt{self-describing \summaries}\xspace}
\newcommand\SDSSFull{\texttt{Self-Describing \Summaries}\xspace}

\newcommand\SDS{\texttt{SDS}\xspace}
\newcommand\SDSS{\texttt{SDS}s\xspace}

\newcommand\SDSscope{data scope\xspace}
\newcommand\SDSlogic{\summary logic\xspace}
\newcommand\SDSpayload{\summary payload\xspace}

\newcommand\createSDS{CreateSDS\xspace}
\newcommand\mergeSDS{MergeSDSs\xspace}
\newcommand\subtractSDS{SubtractSDSs\xspace}
\newcommand\estimateSDS{EstimateFromSDS\xspace}

\newcommand\temporalFiltering{temporal filtering\xspace}
\newcommand\TemporalFiltering{Temporal filtering\xspace}
\newcommand\temporalAggregation{temporal aggregation\xspace}
\newcommand\temporalAggregations{temporal aggregations\xspace}
\newcommand\TemporalAggregation{Temporal aggregation\xspace}
\newcommand\spatialAggregation{spatial aggregation\xspace}
\newcommand\spatialAggregations{spatial aggregations\xspace}
\newcommand\SpatialAggregation{Spatial aggregation\xspace}
\newcommand\spatialFiltering{spatial filtering\xspace}
\newcommand\SpatialFiltering{Spatial filtering\xspace}

\newcommand\datasetAwesomeProm{awesome-prometheus-alerts\xspace}
\newcommand\datasetConviva{prod-enterprise\xspace}
\newcommand\datasetCMU{prod-univ\xspace}
\newcommand\datasetKubernetes{k8s-mixin\xspace}
\newcommand\datasetGrafana{grafana dashboards\xspace]}

\newcommand\qasFull{query-aware \summary\xspace}
\newcommand\qassFull{query-aware \summaries\xspace}
\newcommand\QassFull{Query-aware \summaries\xspace}

\newcommand\qass{\texttt{QAS}es\xspace}
\newcommand\qas{\texttt{QAS}\xspace}

\newcommand\exampleMetric{\texttt{cpu\_usage}\xspace}
\newcommand\exampleQuery{\texttt{avg by (process) (quantile\_over\_time(0.9, cpu\_usage[5m]))}\xspace}

\newcommand\queryWorkload{query workload\xspace}
\newcommand\QueryWorkload{Query workload\xspace}
\newcommand\queryWorkloads{query workloads\xspace}
\newcommand\QueryWorkloads{Query workloads\xspace}

\newcommand\rqeFull{repeating query expression\xspace}
\newcommand\rqesFull{repeating query expressions\xspace}
\newcommand\RQEFull{Repeating query expression\xspace}
\newcommand\RQE{\texttt{RQE}\xspace}
\newcommand\RQES{\texttt{RQEs}\xspace}

\newcommand\aqeFull{repeating atomic query expression\xspace}
\newcommand\AQEFull{Repeating atomic query expression\xspace}
\newcommand\aqesFull{repeating atomic query expressions\xspace}
\newcommand\AQESFull{Repeating atomic query expressions\xspace}
\newcommand\AQE{\texttt{RAQE}\xspace}
\newcommand\AQES{\texttt{RAQEs}\xspace}

\defttself{QE}
\newcommand\QueryExpr{Query expression\xspace}
\newcommand\QueryExprs{Query expressions\xspace}
\newcommand\queryExpr{query expression\xspace}
\newcommand\queryExprs{query expressions\xspace}

\newcommand\Aggregation{Aggregation\xspace}
\newcommand\Aggregations{Aggregations\xspace}
\newcommand\aggregation{aggregation\xspace}
\newcommand\aggregations{aggregations\xspace}

\newcommand\Mv{Materialized view\xspace}
\newcommand\Mvs{Materialized views\xspace}
\newcommand\mv{materialized view\xspace}
\newcommand\mvs{materialized views\xspace}

\deftt{capability}{attribute}
\deftt{capabilities}{attributes}
\deftt{Capabilities}{Attributes}

\newcommand\PerQueryBaseline{PerQuery}
\newcommand\AutoSketchBaseline{AutoSketch}

\newcommand\mergeable{mergeable\xspace}
\newcommand\Mergeable{Mergeable\xspace}
\newcommand\invertible{invertible\xspace}
\newcommand\Invertible{Invertible\xspace}
\newcommand\subpopAware{subpopulation-aware\xspace}
\newcommand\SubpopAware{Subpopulation-aware\xspace}

\newcommand\backwardsCompatible{backwards compatible\xspace}
\newcommand\statusQuoView{query-centric\xspace}
\newcommand\sysnameView{semantic-preserving summary view\xspace}
\newcommand\sps{semantic-preserving summary\xspace}
\newcommand\spss{semantic-preserving summaries\xspace}
\newcommand\reformulationPlan{reformulation plan\xspace}
\newcommand\reformulationPlans{reformulation plans\xspace}
\newcommand\PrecomputeConfig{\texttt{IngestConf}\xspace}
\newcommand\InferenceConfig{\texttt{EstimationConf}\xspace}
\newcommand\SketchOp{\texttt{SketchOp}\xspace}
\newcommand\SketchLib{\texttt{SketchOpLib}\xspace}
\newcommand\NamedSummary{\SDS}
\newcommand{\SimpleTemporal}{\ensuremath{\mathit{OneT}}\xspace}
\newcommand{\SimpleSpatial}{\ensuremath{\mathit{OneS}}\xspace}
\newcommand{\OneSpatialTemporal}{\ensuremath{\mathit{OneSOneT}}\xspace}
\newcommand{\AtomicQuery}{\ensuremath{\mathit{AtomicQuery}}\xspace}

\newif\ifcomments
\commentsfalse

\newcommand\highlight[1]{{{#1}}}

\ifcomments
\newcommand\todo[1]{{\color{red}(\textbf{todo: }{#1}) }}
\newcommand\vyas[1]{{\color{red}\textbf{VS: }{#1} }}
\newcommand\hidden[1]{}
\newcommand\milind[1]{{\color{blue}\textbf{MS: }{#1} }}
\newcommand\alan[1]{{\color{purple}\textbf{AL: }{#1} }}
\newcommand{\zeying}[1]{{\color{orange}Zeying: {#1}}}
\newcommand\story[1]{{\color{brown}{#1} }}

\else
\newcommand\todo[1]{}
\newcommand\hidden[1]{}
\newcommand\vyas[1]{}
\newcommand\milind[1]{}
\newcommand{\zeying}[1]{}
\newcommand\story[1]{}

\fi

\newcommand{\mypara}[1]{\smallskip \noindent{\bf {#1}:}~}
\newcommand{\myparatight}[1]{\noindent{\bf {#1}:}~}

\newcounter{packednmbr}
\newenvironment{packedenumerate}{\begin{list}{\thepackednmbr.}{\usecounter{packednmbr}\setlength{\itemsep}{0.5pt}\addtolength{\labelwidth}{-15pt}\setlength{\leftmargin}{\labelwidth}\setlength{\listparindent}{\parindent}\setlength{\parsep}{1pt}\setlength{\topsep}{0pt}}}{\end{list}}

\newenvironment{packeditemize}{\begin{list}{$\bullet$}{\setlength{\itemsep}{1pt}\addtolength{\labelwidth}{0pt}\setlength{\leftmargin}{\labelwidth}\setlength{\listparindent}{\parindent}\setlength{\parsep}{1pt}\setlength{\topsep}{0pt}}}{\end{list}}

\newcommand{\insightref}[1]{\textbf{Insight~}\ref{#1}}
\newcounter{insightlabel}
\newcounter{insightnmbr}
\renewcommand{\theinsightlabel}{\textbf{\theinsightnmbr}}
\newenvironment{insight}{ 
\begin{list}{\textbf{Insight }\theinsightlabel:~}{\usecounter{insightlabel}\stepcounter{insightnmbr}\setlength{\labelwidth}{0pt}\setlength
{\labelsep}{0pt}\setlength{\leftmargin}{0in}\noindent\rule{\linewidth}{0.5pt}\vspace{-10pt}\item \bf \em}}{\\*[-7pt]\end{list}\vspace{-10pt}\noindent\rule{\linewidth}{0.5pt}}

\newcommand{\challengeref}[1]{\textbf{Challenge~}\ref{#1}}
\newcounter{challengelabel}
\newcounter{challengenmbr}
\renewcommand{\thechallengelabel}{\textbf{\thechallengenmbr}}
\newenvironment{challenge}{ 
\begin{list}{\textbf{Challenge }\thechallengelabel:~}{\usecounter{challengelabel}\stepcounter{challengenmbr}\setlength{\labelwidth}{0pt}\setlength
{\labelsep}{0pt}\setlength{\leftmargin}{0in}\noindent\rule{\linewidth}{0.5pt}\vspace{-10pt}\item \bf \em}}{\\*[-7pt]\end{list}\vspace{-10pt}\noindent\rule{\linewidth}{0.5pt}}

\newcommand{\hlfunc}[1]{\textcolor{functioncolor}{\textbf{#1}}}

\newcommand{\hlarrow}{\textcolor{arrowcolor}{\textbf{->}}}

\newcommand{\findingref}[1]{\textbf{Finding~}\ref{#1}}
\newcounter{findinglabel}
\newcounter{findingnmbr}
\renewcommand{\thefindinglabel}{\textbf{\thefindingnmbr}}
\newenvironment{finding}{ 
\begin{list}{\textbf{Finding }\thefindinglabel:~}{\usecounter{findinglabel}\stepcounter{findingnmbr}\setlength{\labelwidth}{0pt}\setlength
{\labelsep}{0pt}\setlength{\leftmargin}{0in}\noindent\rule{\linewidth}{0.5pt}\vspace{-10pt}\item \bf \em}}{\\*[-7pt]\end{list}\vspace{-10pt}\noindent\rule{\linewidth}{0.5pt}}

\newcommand{\queryBox}[1]{%
  \tcbox[on line, nobeforeafter,
         colback=gray!8, colframe=gray!60,
         boxrule=0.4pt, arc=1.5pt,
         left=2pt, right=2pt, top=1pt, bottom=1pt,
         boxsep=0pt]{\texttt{\small\detokenize{#1}}}%
}

\lstdefinelanguage{PromQL}{
  morekeywords={topk, bottomk, quantile_over_time, avg_over_time,
                rate, sum, by, without},
  sensitive=true,
  morestring=[b]",
}

\lstdefinestyle{query}{
  language=PromQL,
  basicstyle=\ttfamily\small,
  keywordstyle=\color{blue!70!black},
  stringstyle=\color{red!70!black},
  backgroundcolor=\color{gray!6},
  frame=single, rulecolor=\color{gray!60},
  xleftmargin=4pt, xrightmargin=4pt,
  framexleftmargin=2pt,
  columns=fullflexible, keepspaces=true,
  breaklines=true, breakindent=2em,   %
  showstringspaces=false,
  aboveskip=6pt, belowskip=6pt,
}

\maketitle

\begin{abstract}

Many large enterprise  and cloud deployments rely on {\em metrics observability} solutions (e.g., Prometheus) to monitor key performance indicators of system health (e.g., CPU, memory, network, app-level statistics). 
 Monitoring and analyzing these metrics    presents a significant challenge in terms of cost-latency tradeoffs at scale. 
 Ideally, system operators would like {\em  immediately deployable and backwards compatible} solutions to improve the {\em cost-latency}  tradeoffs without significant impact on the {\em fidelity} of their  observability applications. 
  Unfortunately, existing solutions including state-of-art timeseries databases and custom approximate querying   solutions fail on one or more aspects.

 In this paper, 
 we  present the design and implementation of \sysname,  a backwards-compatible drop-in accelerator for metrics observability deployments.  
 \sysname recasts metrics analytics tasks  using
a \textit{\sysnameView}   rather than classical   \statusQuoView view. Instead of estimating query results on raw timeseries metrics data,  \sysname uses \spss  that enable high-fidelity analytics with significant cost-latency improvements.
We present a practical implementation of \sysname that can  be retrofitted to  {\em any} existing observability deployment. 
We evaluate \sysname on diverse real and synthetic workloads against state-of-art solutions and show that it can   provide  significant  reduction in query latency and  cost. \sysname is open-source at \href{https://github.com/ProjectASAP/ASAPQuery}{https://github.com/ProjectASAP/ASAPQuery}.

\end{abstract}

\section{Introduction}

\begin{figure}[t]
    \centering
    \includegraphics[width=\linewidth]{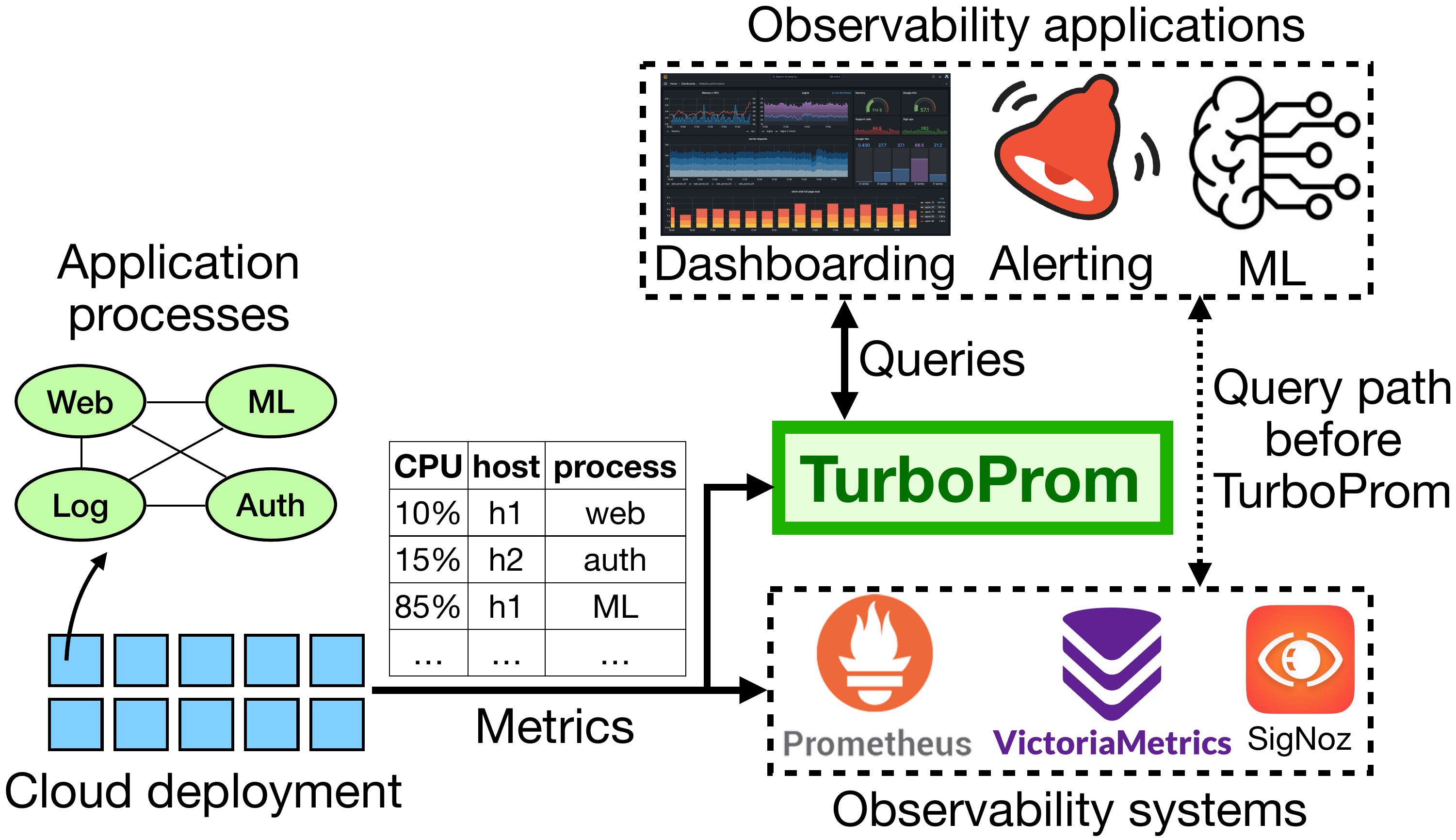}
    \caption{\sysname is backwards-compatible with an existing \mo deployment.}
    \label{fig:canonical_olly_deployment}
\end{figure}

Large enterprise and cloud deployments rely on {\em \molong} systems to  monitor and analyze  the health of their  computing    infrastructure. Prometheus  is a  popular  de-facto metrics observability  system~\cite{prometheus} with widespread adoption.  Operators instrument their systems to export key  {\em metrics} (e.g., CPU, memory, network, app-level metrics) along with {\em labels} (e.g., services, locations, versions) to a \molong (\mo in short) system (typically backed by a  timeseries database~\cite{prometheus-tsdb, signoz-clickhouse}). They  create  custom dashboards, alert rules, and run  real-time anomaly detection algorithms atop such systems.

As systems grow in scale and complexity, the cost, scale, and responsiveness needs of  \mo  also grow commensurately, as  documented by many industry  reports (e.g.,~\cite{high-obs-cost-industry-1, high-obs-cost-industry-2, high-obs-cost-industry-3}).
For instance,  Uber~\cite{uber-blog-1}  reports serving 2500 queries on top of 8.5 billion metrics each second.
In response, the  academic and industrial community have crafted many     solutions to tackle this challenge.

Ideally, we want solutions that (R1) are {\em immediately deployable}   and {\em \backwardsCompatible} with the existing  infrastructure;  (R2)   can  significantly improve the {\em cost-latency tradeoff  at scale}; and (R3) minimally  impact  the {\em analytics fidelity}.
While existing solutions have made great inroads, they fail to simultaneously satisfy R1--R3. 
For instance,  other  timeseries databases~\cite{clickhouse}  satisfy R3, but are not \backwardsCompatible and only provide limited benefits for R2. Other  designs  address R2 for   specific queries (e.g., quantiles)~\cite{bigquery-sketches}, but sacrifice R1. Some other solutions use  coarse sampling or  drop  data  that can maintain R1 and R2, at the cost of R3~\cite{strawman-approach-1, strawman-approach-2, strawman-approach-3, strawman-approach-4}.

 We take a first-principles approach to   understand  the needs of \mo analytics to systematically  tackle   R1--R3. We analyze the structure of \mo \queryWorkloads across multiple settings and observe that: (1) these queries are {\em amenable to compact \spss}; (2) there is significant {\em query predictability} across time; and  
(3) there is significant {\em duplicate computation} in workload across queries and across time.

Based on these insights, we recast  query  estimation in \mo   using a \textit{\sysnameView}. 
 Instead of computing  the \queryWorkloads as-is on raw data, we argue that  the result of the query can be estimated using  semantically-equivalent  operations on summarized data structures. Let us consider an example.  Suppose   the  operator wants to check whether ``most'' machines in the cluster had ``high'' utilization recently. Today,  running a query for this intent will   scan all  raw metrics  data   in  the  timeseries database to  compute  median CPU utilization  for  each   machine  over the last 15 mins,  and then find  the global average.    However, if we could  compute the median and average without scanning all the raw data (e.g., using data streaming algorithms using highly efficient approximate algorithms like   sketches~\cite{kll}), that could drastically reduce cost and latency.

To translate this high-level idea into practice, we design and implement  \sysname,  a \backwardsCompatible drop-in accelerator for \mo deployments.
 \sysname uses an offline \Planner that runs a profiling step to understand the data and query model of a given deployment, and creates a custom \reformulationPlan. This  identifies and exploits opportunities for using compact summaries,  ingest-time precomputation, and reducing duplicate computation.
  \sysname's  \Summarizer \textit{precomputes} compact \spss on raw \metrics data as it is streamed in, and periodically exports  summaries.   
 The \QueryEngine \textit{intercepts and accelerates queries} before they hit the existing \mo system.   It  uses the computed summaries to estimate results. Unoptimizable queries fall back to the legacy \mo system.

  In designing \sysname, we address key technical challenges.  
 First, we introduce a novel abstraction of a \NamedSummary (\SDS) that represents an atomic unit of summarization over a metrics data stream  (e.g., a  window average for a specific label filter for a specified time window).   %
 Second, to tackle the modeling and combinatorial explosion  in the planning problem,  we design and implement a  simple-yet-effective heuristic algorithm in the \Planner that takes a historical \queryWorkload, reformulates it to use \SDSS and computes effective summary-aware \reformulationPlans. 
  Third, we implement an efficient streaming  library of  compact sketch-based summaries that covers  many common queries  in the \Summarizer to produce an ensemble of \SDSS.  
 Finally, we  design an efficient \QueryEngine that can estimate  queries using  the precomputed \SDSS if an optimized estimation plan is available, and falls back to the legacy \mo system otherwise.

 Our  \sysname implementation is  agnostic to the specific \mo  deployment,  as long as it is consistent with the Prometheus data and query model.  We implement Python modules for the offline tasks, extend an open-source Rust-based streaming~\cite{arroyo} for the \Summarizer, and implement a custom efficient \QueryEngine in Rust.
 We evaluate \sysname using a combination of public and synthetic workloads against state-of-the-art MO systems like \Prom and \VM. Across diverse workloads at small scales, we find that \sysname reduces query cost and latency over baselines by 2x and 10x, respectively. For different query types in isolation, \sysname's benefits increase with query load, providing up-to 3800X latency reduction. Extrapolating these benefits to various workloads at hyper-scale, we estimate  \sysname can provide 1.2X--900X benefits across an entire workload.

\section{Background on \molong (\mo)}
\label{sec:background}

\Mo (\mo) is the instrumentation, collection, storage, and analysis of numerical telemetry data from a software deployment. Operators use this telemetry to enable real-time downstream tasks such as alerting~\cite{alering-rules}, anomaly detection~\cite{downstream-task-anomaly-1, downstream-task-anomaly-2, downstream-task-anomaly-3}, autoscaling~\cite{downstream-task-autoscaling}, and in CI/CD pipelines~\cite{downstream-task-canary-deployment}. 
For instance, a Kubernetes cluster operator may want to monitor the CPU and memory usage of each pod. The de-facto standard for \mo is  \Prom,  an open source system~\cite{prometheus}. \Prom supports  a wide range of applications such as databases and web servers~\cite{prometheus-exporters}. Other popular systems for \mo include \VM~\cite{victoria-metrics} and \Clickhouse~\cite{clickhouse}.

\Mo \metrics are time-series data associated with metadata. Each \metric instance has a metric \textit{name}, numerical \textit{measurement}, a \textit{timestamp}, and a set of key-value \textit{metadata labels}. For e.g. the instantaneous CPU usage of a docker container can be represented with \textit{value}=\% CPU usage and \textit{labels} representing the docker container ID, service name, and host name. These \metrics are usually recorded at a fixed time interval. An \mo query comprises of filtering, aggregation, and binary arithmetic operators. Both filtering and aggregation operators can be either temporal or spatial, operating on timestamps and metadata labels, respectively. Binary operators like $+,-,*,/$ combine results of two queries.

\subsection{Motivating use-case scenarios}
\label{subsec:mo_workload_structure}

Let us now consider a few motivating \uc scenarios that drive cost-latency problems in \mo deployments.

\todo{frame as operational questions for each scenario}

\mypara{Performance monitoring}
Operators continuously monitor the performance of their software deployments, such as request latencies, resource utilization and SLO violations and identify poorly performing components~\cite{google_sre,datadog_slo,datadog_cpu}. For e.g., operators may want to rank services by tail latency, identify pods experiencing sustained resource saturation or identify services with excessive error rates.

\begin{lstlisting}[style=query]
topk(10, histogram_quantile(0.95,
  sum by(service,le)(rate(latency_bucket[5m]))))
  
quantile_over_time(0.9, cpu_by_pod[1h]) > 0.8

sum by(service)(rate(errors_total[5m])) /
  sum by(service)(rate(requests_total[5m])) > 0.01
\end{lstlisting}

\mypara{Anomaly detection}
Operators detect deviations and anomalies by comparing distributional measures over different data-subpopulations, and against historical behavior. For e.g., Grafana uses rolling averages and standard deviations to construct anomaly bounds, while LinkedIn's ThirdEye dynamically selects high-contribution subpopulations for monitoring \cite{grafana_anomaly,linkedin_thirdeye}. Elastic provides mechanisms for population analysis~\cite{elastic_population_analysis} to detect entities deviating from their peers.

\begin{lstlisting}[style=query]
abs(x - avg_over_time(x[1h])) > 3 * stddev_over_time(x[26h])

topk(10, abs(traffic - avg_over_time(traffic[1d])) and (traffic > 100))

topk(10, stddev by(cluster)(cpu_by_node))
\end{lstlisting}

\mypara{Trend analysis}
Operators periodically analyze resource consumption to identify inefficient allocations, growing populations, and capacity pressure. For e.g. Datadog and Grafana describe how track disk exhaustion, CPU utilization and time-series cardinality across applications~\cite{datadog_forecast,datadog_cpu,datadog_alerting101,grafana_cardinality}. These workloads typically use longer windows and are re-evaluated less frequently (e.g., every few hours).

\begin{lstlisting}[style=query]
topk(10, cpu_requested - avg_over_time(cpu_used[7d]))

topk(10, count by(namespace)(kube_pod_info) - avg_over_time(
    (count by(namespace)(kube_pod_info))[1d:5m]))

topk(10, sum by(cluster)(disk_used_bytes) - avg_over_time(
    (sum by(cluster)(disk_used_bytes))[7d:1h]))
\end{lstlisting}

\mypara{Unique structure of \mo workloads}
From these motivating scenarios, we see that operators deploy an ensemble of queries together to serve a particular downstream task. These queries repeat continuously over time, are not single query instances that are issued in an ad-hoc manner.
Each query instance itself computes one or more statistical aggregations (sum, count, avg, topk, quantile, etc.) that can aggregate values across a time range, or across metadata labels, or both. For e.g. \queryBox{quantile_over_time} computes quantiles across a time range, for each metadata label combination while \queryBox{quantile by(cluster)} computes quantiles for each value of the cluster label\footnote{Similar to a SQL query ``GROUP BY cluster''}. Thus, these aggregations are computed over different sub-populations of the underlying high-dimensional \metrics data. Putting all this together, we observe that \mo tasks exhibit a unique \queryWorkload structure -- they are an \textit{ensemble} of real-time \textit{repeating spatio-temporal aggregations} on \textit{time-series data}.

\subsection{Problem: Cost, scale, and latency}

\begin{figure}[thb]
    \centering
    \includegraphics[width=220pt]{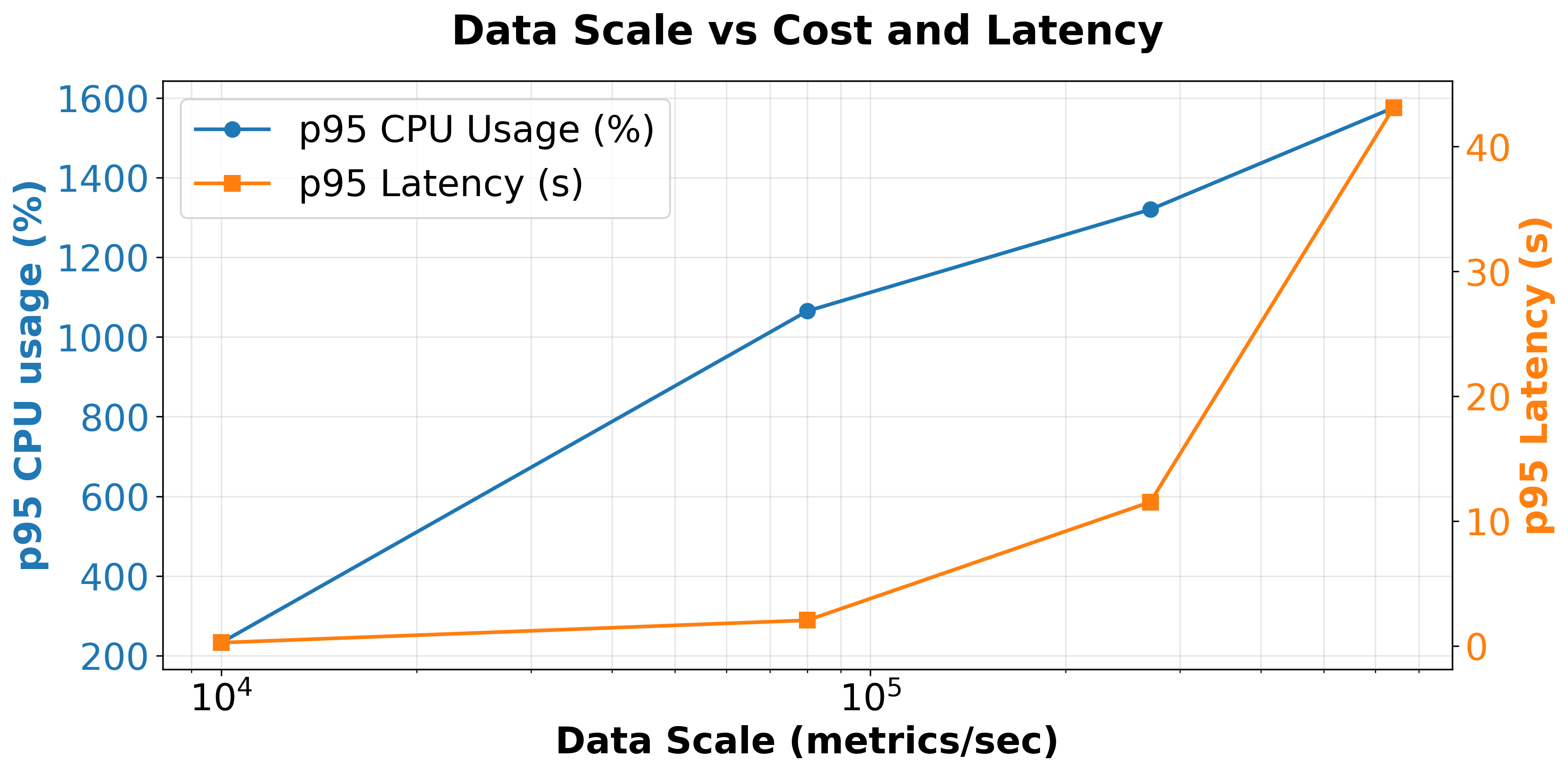}
    \vspace{-0.3cm}
    \caption{As deployments scale up, the cost and latency of \mo queries rises untenably.}
     \vspace{-0.3cm}
    \label{fig:observability_cost}
\end{figure}

Operators running \mo deployments for the above downstream \ucs care about two measures of success: Total Cost of Ownership (TCO), and query latency.
(1) The TCO measures the entire cost of operating the \mo infrastructure. In this work, we consider the self-hosted model where operators deploy and manage \Prom in a cloud such as AWS. Here, TCO equals the cost of the provisioned machines, which depends on the CPU and memory requirements of \Prom. TCO includes cost for ingesting \metrics and running \queries. 
(2) Query latency measures time taken between a query being issued and its result being ready. Operators would like to monitor large-scale deployments with low TCO and latency. 

As cloud deployments scale up, the \mo cost and latency to serve various \ucs rises. To quantify these effects, we ran 10 \PromQL queries every second against a \Prom instance on a 16-core 2.0 GHz machine. We vary the offered load and measure the p95 query latency and p95 CPU usage. 
In Figure~\ref{fig:observability_cost}, we see that at  an offered load of  640k metrics/sec, the p95 query latency  is 40 seconds and p95 CPU usage of 14  cores.
Suppose the operators have an SLO of 5 sec p95 latency. Even assuming  perfect scalability, this would mean that handling  a Uber-like \queryWorkload~\cite{uber-blog-1}  with 2500 queries/sec would need  $(2500/10)\times(40/5)$ = 2000 machines. Using AWS EC2's ``c8g.4xlarge'' (a compute-optimized instance with 16 vCPUs), we estimate this would cost  $\approx$ \$448,000 a month to support this workload! Indeed, multiple enterprises have documented exorbitantly high costs for their \mo deployments~\cite{high-obs-cost-industry-1, high-obs-cost-industry-2}.

\section{Overview}
\label{sec:overview}

\subsection{High-level idea}
\label{subsec:high_level_idea}

\begin{figure}[t]
    \centering
    \includegraphics[width=\linewidth]{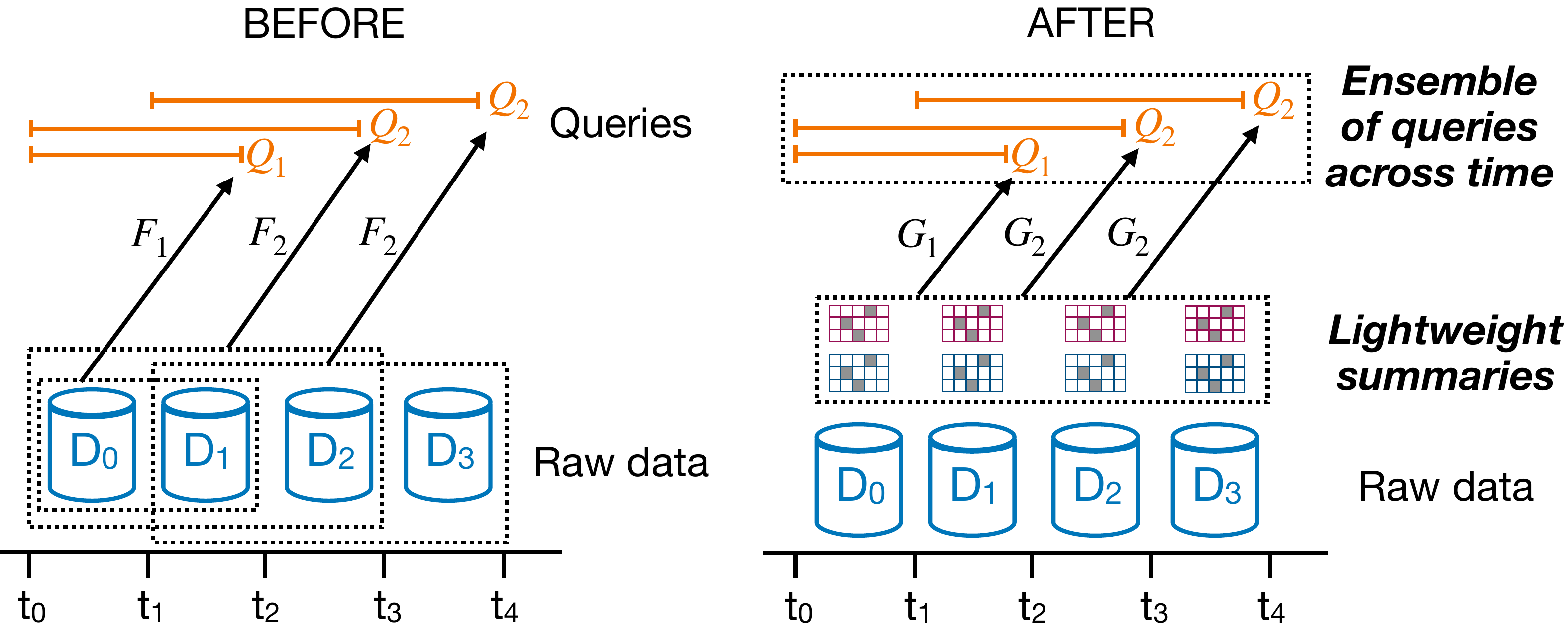}
    \vspace{-0.3cm}
    \caption{\sysname reformulates \mo \queryWorkloads to use \summary-based \mvs}
    \vspace{-0.3cm}
    \label{fig:design_insight}
\end{figure}

\sysname's key idea is to automatically create and use \summary-based \mvs for \mo \queryWorkloads. Given an \textit{ensemble of repeating queries} with spatio-temporal aggregations, we automatically (a) construct \summary-based \mvs that operate on time-series \metrics data, and (b) re-write queries to use materialized \summaries to reduce total cost and improve query performance. Each \summary may be approximate (e.g. KLL sketch~\cite{kll}) or an exact representation of the underlying data. Figure~\ref{fig:design_insight} illustrates this idea. The left half depicts traditional \mo where each query is executed in isolation on top of raw data. The right half depicts \sysname's idea where we automatically materialized \summaries and then map the repeating query ensemble to execute on top of the \summaries.

\Summary-based \mvs enable three complementary opportunities for \mo \queryWorkloads.

\mypara{Reuse across queries}
\Summary-based \mvs allow us to reduce duplicate computation across queries. Consider two queries: \texttt{quantile by () (0.5, cpu\_usage)} and \texttt{quantile by () (0.9, cpu\_usage)} that compute the median and p90 CPU usage. While it seems like these queries do not have common computation that can reused, if we rewrite the query to use a KLL sketch, we can estimate both median and p90 from the same sketch \summary.

\mypara{Reuse across time}
\Summary-based \mvs also allow us to reduce duplicate computation that occurs from repeating instances of a query. Consider the query \texttt{quantile\_over\_time(0.5, cpu\_usage[5m])} that repeats every 30 sec. If each instance of this sliding window query is executed individually, each data point is re-processed by 10 query windows (5 min / 30 sec = 10) resulting in unnecessary compute. Instead, we can compute intermediate \qasFull on disjoint 30 sec chunks of data, and then merge the appropriate \summaries to estimate the result of each instance of the sliding window query.

\mypara{Reducing query latency}
Executing queries over \summaries instead of raw data allow us to break the query execution into two parts, since the computation of a \summary from raw data can be decoupled from the estimation of a query result from the \summary. This is useful as we can leverage the predictable nature of \mo's \queryWorkloads to compute \summaries at ingest-time as data is streaming in, and estimate results from \summaries when queries are actually issued. Essentially, this enables us to move some compute from query-time to ingest-time, thus reducing query latency.

\subsection{Challenges}
\label{subsec:challenges}

Automatically creating \summary-based \mvs and re-writing queries to use them introduce three research challenges.

\begin{challenge}{
Defining an abstraction for materialized \summary instances
\label{challenge:abstraction}
}
\end{challenge}

\todo{challenge is not coming across. Narrow waist has to be up-leveled}

The first challenge is an abstraction challenge -- how do we represent materialized \summaries in practice? What is the data structure we should use?
A strawman solution is to use abstractions from existing systems. \Mo systems like \Prom~\cite{prometheus} and \VM~\cite{victoria-metrics} use a time-series abstraction where an atomic piece of data is a metric instance that is associated with a name, timestamp, value and a set of labels. Databases represent every piece of data as a table, wherein each table has a defined schema. Spark~\cite{spark} introduced a new abstraction of  resilient distributed datasets (RDDs) to make fault tolerance easier.
Unfortunately, these abstractions are ill-suited because a materialized \summary is neither a time-series of raw measurements, nor a SQL table, nor an RDD. Instead, it is a different type of object that connects queries to metrics data. Thus, we must define a new abstraction to handle it.

This is challenging because the abstraction must be a narrow-waist between a diverse set of queries (the entire space of possible \PromQL) and a diverse set of \summary algorithms (such as CountMinSketch~\cite{cmsketch}, KLL~\cite{kll}, HLL~\cite{hll}, DDSketch~\cite{ddsketch}, etc.). This abstraction must be extensible to support more queries (for e.g. \VM's MetricsQL~\cite{metricsql} extends \PromQL~\cite{promql}) and newer \summary algorithms (e.g.~\cite{omnisketch, micromon, sublime_sketch}.

\begin{challenge}{
Automatically deciding what \summary-based \mvs to create and how to re-writing queries to use them
\label{challenge:mapping}
}
\end{challenge}

The second challenge is an algorithmic one -- given a \mo \queryWorkload in a high-level language such as \PromQL or SQL, how do we automatically infer the optimal set of \summary-based \mvs to create and re-write queries to use them? This has two challenges. 
(1) First, any \mo query is an arbitrary combination of spatio-temporal filtering and aggregation operators, and binary operators (as we showed in \S\ref{subsec:mo_workload_structure}). A \queryWorkload is an arbitrary combination of such repeating  queries, and thus the space of possible \queryWorkloads is the powerset of the set of the possible queries and is intractable to reason about.
(2) Second, for any given \mo query, we do not know how to automatically construct \summary-based \mvs that can feasibly estimate the query.  While there is extensive amount of literature on query planning~\cite{query-plan-1, query-plan-2, query-plan-3, query-plan-4}, these tackle the problem of executing queries on raw data, not on reformulating a repeating ensemble of time-series queries to execute on \summaries.

\begin{challenge}{
Designing an efficient system to realize \summary-based materialization
\label{challenge:realization}
}
\end{challenge}

Finally, we need a system that can realize these concepts in practice. This system must be able to implement a large space of \summary-based \mvs and map high-level queries to use these summaries.
This requires an efficient data plane that continuously ingests new \metrics data, and computes and materializes \summaries on it, and then uses them to estimate an ensemble of results for the repeating \queryWorkload . At the same time, this system must integrate easily within an existing \mo deployment, to reduce deployment friction and drive adoption. A strawman solution is to modify an existing system like \Prom, \VM or \Clickhouse but this is untenable as \sysname's view of \summary-based materialization as a first-class principle is at odds with the abstractions and design principles of these systems. Thus, we design \sysname from scratch.

\subsection{System overview}
\label{subsec:overvire}

\begin{figure}[t]
    \centering
        \includegraphics[width=\linewidth]{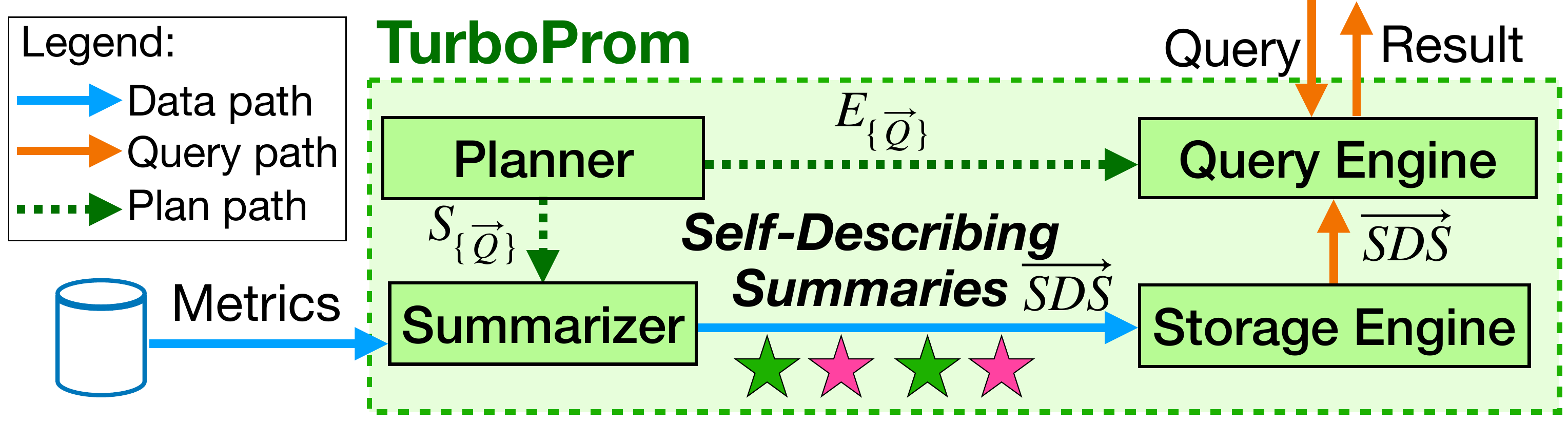}
        \caption{Overview of \sysname's design. The control plane consists of the \Planner and decides the \qassFull ($\qasVectorMath$) to compute ($\summarizeFuncMath$) and how to estimate queries from them $\estimateFuncMath$. The data plane realizes these functions and stores materialized $\qasVectorMath$.}
        \vspace{-0.3cm}
    \label{fig:system_overview}
\end{figure}

\sysname is a system to accelerate analytics and reduce the cost of \mo. We design \sysname as a drop-in system that integrates into an existing \mo deployment, to reduce friction in adoption and enable incremental deployment. 
In designing \sysname, we make the following core contributions that directly tackle the challenges from \S\ref{subsec:challenges}:
\begin{enumerate}
    \item We design a novel abstraction to represent materialized \summaries: \SDSSFull (\SDSS). 
    \item We design a planner that automatically infers the optimal set of \summary-based \mvs to create from a high-level \queryWorkload in SQL or \PromQL.
    \item We design an efficient systems to compute and materialized \SDSS from raw metrics, and use them to estimate results for an ensemble of repeating \mo queries.
\end{enumerate}

\sysname has four main components, shown in Figure~\ref{fig:system_overview}.
\sysname's control plane consists of the \Planner that takes a high-level query workload as input from the user and automatically reformulates it to use \SDSS, based on the high-level idea from \S\ref{subsec:high_level_idea}. This reformulation is described by (a) what \SDSS to materialize, and (b) how to re-write queries to use these \SDSS.
The user may optionally specify an accuracy requirement for each query provided to the \Planner.
Since \mo query workloads change infrequently, we envision the \Planner running as an infrequent offline step that can reformulate the plan whenever the data or query workload changes (say, every week or so).

\sysname's data plane consists of the \Summarizer, the \StorageEngine and the \QueryEngine. As new data is collected, it is routed to the \Summarizer which computes \SDSS in a streaming manner, based on the \Planner's output. Materialized \SDSS are written to the \StorageEngine which persists them to disk and maintains a cache of recent \SDSS in memory.
When a batch of queries is issued from a \mo dashboard or query client, these queries are routed to the \QueryEngine. The \QueryEngine reads the relevant \SDSS, manipulates them and estimates query results, as dictated by the \Planner.

Some queries may be too complex for the \QueryEngine to estimate, may not map to a corresponding \SDS due to being issued in an ad-hoc manner or may simply be queries without aggregation that are already cheap to compute. In such cases, the \QueryEngine falls back to the existing \mo backend. As a result, any \mo query workload always has 100\% query coverage and is no worse off than before as every query always has a fallback option.

\section{Detailed Design of \sysname}
\label{sec:design}

We start by describing \sysname's core abstraction of ``\SDSSFull''. We then describe its control plane, consisting of the \Planner module. Finally, we discuss the data plane that comprises the \Summarizer, \StorageEngine, and \QueryEngine.

\subsection{Core abstraction: \SDSFull}
\label{subsec:sds}

\sysname's idea is to compute an ensemble of \qassFull  from raw \metrics time-series $\dataVectorMath$ and use the computed \summaries to estimate an ensemble of repeating \mo queries.
To practically realize the \qassFull concept, we require a ``narrow-waist'' abstraction that unifies the set of possible \summaries and is extensible to support new queries and summary algorithms. Based on this, we design \sysname's core abstraction: \SDSFull (\SDS).

The intuition is that any \SDS must describe \textit{what} was summarized, \textit{how} it was summarized, and the resulting \summary.
Thus, we define each \SDS as being composed of these three parts: the \textit{\SDSscope}, \textit{\SDSlogic}, and \textit{\SDSpayload}.

The \textit{\SDSscope} describes \textit{what} was summarized. It is specified by the name of the summarized metric, the time range of data that was summarized, an optional set of filters applied on metadata labels before summarization, and, an optional set of metadata labels used to partition the data before summarization.
The \textit{\SDSlogic} describes \textit{how} the data was summarized. It is specified by a unique name of the \summary algorithms (e.g. CountMinSketch, KLL, DDSketch, Hydra, etc) and its associated configuration. The configuration depends on the exact algorithm. For instance, a KLL sketch only takes a single configuration parameter $k$ that denotes the number of compactors. A CountMinSketch takes two configuration parameters denoting the number of rows and columns in the counter array. A more complicated sketch like Hydra~\cite{hydra} takes multiple parameters: a list of label columns used for hashing into the outer sketch, number of rows and columns of the outer sketch, and the rows, columns and depth of the inner sketch.
Finally, the \textit{\SDSpayload} contains the actual serialized bytes representing the materialized \summary output. This representation, too, depends on the exact algorithm.

\begin{figure}[t]
    \centering
    \includegraphics[width=\linewidth]{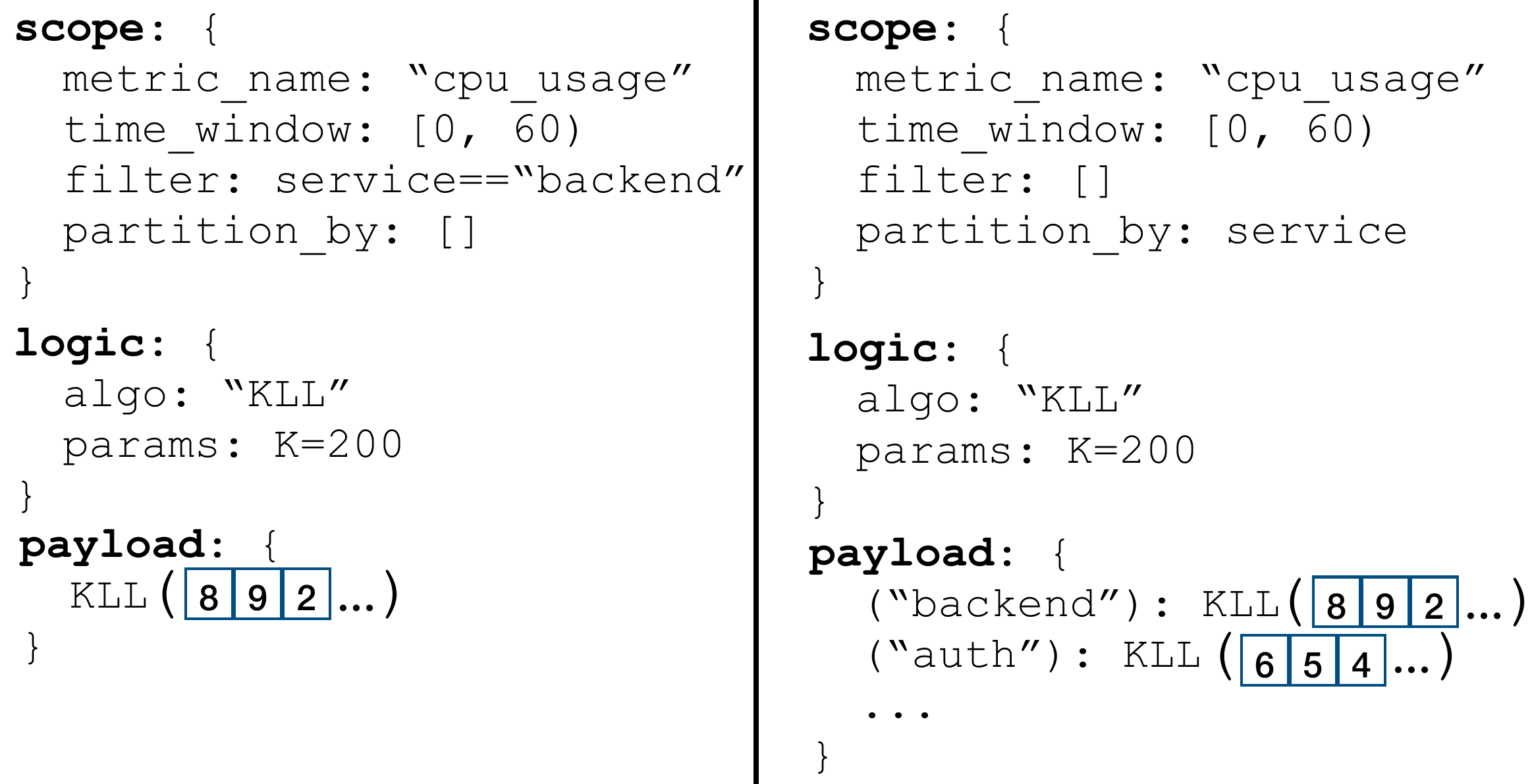}
    \caption{Example \SDSSFull}
    \label{fig:sds_example}
\end{figure}

Let us illustrate these concepts with two examples shown in Figure~\ref{fig:sds_example}. Consider an example metric \exampleMetric with labels \texttt{host} and \texttt{service} that represents the CPU usage of different Docker containers across hosts running services like ``frontend'', ``auth-server'' and ``backend''.
An operator may want to understand the p90 CPU usage for the time range $[0, 60]$ for the backend service. The first \SDS can help them do this.
Alternatively, the operator may want to understand p90 CPPU for the same time range, but for each of the three services -- the second \SDS contains multiple KLL sketch-based \summaries computed on data partitioned by the service label.

The \SDSscope and \SDSlogic allow a consumer to determine whether an \SDS it can be used to answer a query, without accessing the raw data used to construct the \SDS. Further, the \SDS abstraction allows \sysname to easily extend to new queries and \summary algorithms.

\subsection{\sysname's control plane: \Planner}

\todo{up-level \summary-first materialized views across the repeating query workload}

\sysname's control plane consists of the \Planner module and is responsible for analyzing and reformulating a \mo \queryWorkload to run on an ensemble of \sdssFull ($\sdsVectorMath$).
Recall from \S\ref{subsec:mo_workload_structure} that \mo \queryWorkloads are ensembles of repeating queries. Each query is not just a single isolated instance but repeats at fixed intervals, and is part of a larger set of queries issued by a \mo application such as a monitoring dashboard.

\mypara{Inputs and outputs}
Given this workload structure, the \Planner takes two inputs. (1) First, a set of \rqesFull (\RQES) that represent the \mo \queryWorkload, denoted as $\RQESetMath = \{\RQEMath\}$. Each \RQE $\RQEMath$ is a PromQL query expression associated with a fixed time interval of repetition. The operator may specify accuracy and query latency constraints for each \RQE.
(2) Second, a library of implemented \summary algorithms i.e. \SummaryLib = $\{\texttt{\SummaryOp}\}$. Each \SummaryOp may be approximate (e.g. KLL~\cite{kll}) or exact.
The output of the \Planner is a set of summarization plans $\summarizeFuncMath$ and a set of estimation plans $\estimateFuncMath$, and a mapping from \RQE to a specific summarization-estimation plan pair $(\summaryChainMath,\estimateDAGMath)$. Here, $\summaryChainMath \in \summarizeFuncMath$ and $\estimateDAGMath \in \estimateFuncMath$. 
The summarization plans describe the set of \SDSS to compute from raw metrics, while estimation plans describe how to estimate query results for \RQES using the computed \SDSS. The objective is to find the set of summarization and estimations plans that minimize the total ingestion and query cost
, while satisfying the accuracy and latency requirements of each \RQE. 

Operators care about two metrics of success. (1) The \Planner must be \textit{effective} i.e. it must produce plans that are near-optimal and improve the cost-performance tradeoff of the input \RQES significantly. (2) The \Planner must be \textit{efficient} i.e. it must reasonably fast in analyzing the input \queryWorkload and producing plans.

One strawman solution is to do a brute-force enumeration of all possible plans, and then select the optimal plans. Even if a brute-force enumeration were possible\footnote{We show later how the space of possible plans is infinite.}, this approach is not \textit{efficient}. Another approach is to support only simple queries that we can directly map to a \SummaryOp. For e.g. we can manually map the query \texttt{quantile(0.5, cpu\_usage)} to a KLL \summary~\cite{kll}. This approach is not \textit{effective} since \mo \queryWorkloads consist of \rqesFull, wherein each query expression is an arbitrary combination of spatio-temporal filtering and aggregation operations.

\mypara{Challenges}
Designing an effective and efficient \Planner is hard because of two key reasons:
(1) First, we do not know how to reformulate a set of \RQES to use \summary algorithms and run on \SDS instances. 
Any query expression in an \RQE is a combination of spatio-temporal filtering and aggregation operations, and thus can be arbitrarily complex. We do not know how to map such complex repeating query expressions to different \SummaryOp and estimate query results.
Existing query planning literature~\cite{query-plan-1, query-plan-2, query-plan-3, query-plan-4} tackles the problem of executing queries on raw data, not on reformulating an ensemble of repeating queries to run on \SDSS.
(2) Even if we tackle this challenge, there is a combinatorial explosion in mapping an \textit{ensemble} of \RQES to an \textit{ensemble} of summarization and estimation plans. The space of \RQE ensembles is the powerset of all possible \RQES. To compound the challenge, the space of possible \SDSS, and thus the space of possible summarization and estimation plans is too large and intractable to reason about. The \SummaryLib consists of multiple \SummaryOp, each of which can be used to create arbitrary \SDSS with different \textit{\SDSscope{s}} and \textit{\SDSlogic{s}}. 
Given this large space of possible plans, it is intractable to  find the optimal set of plans for a given ensemble of \RQES.

\begin{figure}[t]
    \centering
    \includegraphics[width=0.6\linewidth]{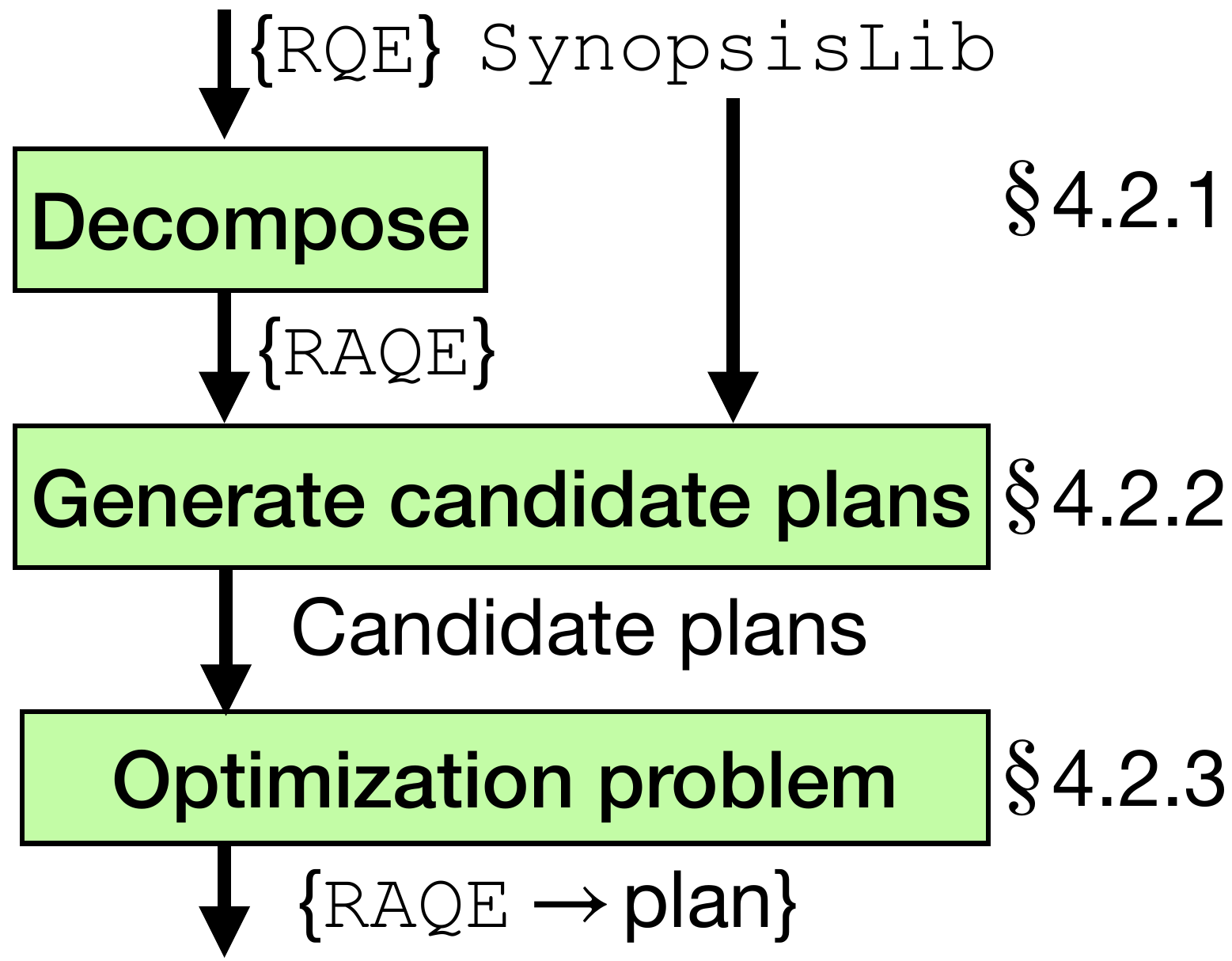}
    \caption{\sysname's \Planner}
    \label{fig:planner_design}
\end{figure}

\mypara{High-level approach}
Figure~\ref{fig:planner_design} shows our high-level approach and design choices to make this problem tractable.
(1) Instead of reasoning about arbitrarily complex \RQES to find the optimal set of plans, we decompose each \RQE into simpler atomic units -- \AQESFull (\AQES) -- that are easier to reason about. Each \AQE represents a subset of the \RQE that does not have any binary operators $+,-,*,/$, scalars or nested aggregations. This is safe since inferring the result of an \RQE from the results of its constituent \AQES is trivial. 
(2) Instead of optimizing over an infinite space\footnote{Plans can summarize arbitrarily small time ranges of data and thus the space of plans is infinite.} of summarization and estimation plans, we formulate the optimization problem to operate over a discrete set of possible plans. To generate this discrete set, we taxonomize the \SummaryLib library and generate candidate plans that are ``feasible-by-construction''.
(3) Finally, we formulate and solve an optimization problem that selects the set of summarization and estimation plans to deploy that minimize total cost and respect latency and accuracy requirements. We use an offline benchmark and an analytical cost model to estimate the costs and accuracy of different plans.

\subsubsection{Decomposing \RQES into \AQES}
\label{subsubsec:decompose}

\begin{figure}[t]
    \centering
    \includegraphics[width=0.7\linewidth]{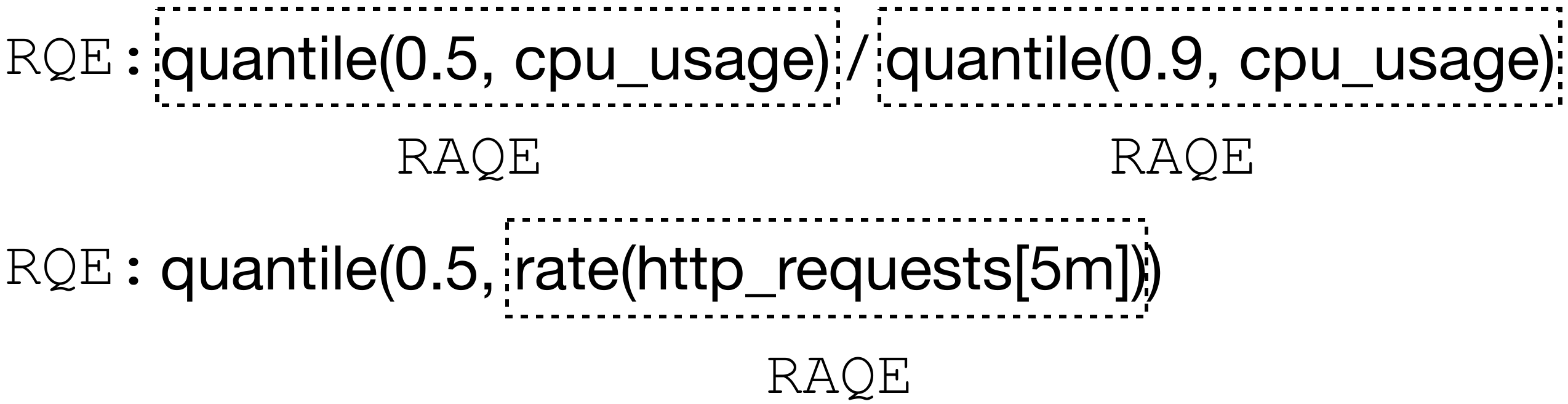}
    \caption{Examples of decomposing an \RQE into \AQES.}
    \label{fig:example_decomposition}
\end{figure}

The aim of decomposing an \RQE into constituent \AQES is to make it easier to reason about reformulating a \queryWorkload to execute using \summaries. We define each \AQE to be a query expression with only a single aggregation, without multiple nested aggregations and without any binary operators $+,-,*,/$. Figure~\ref{fig:example_decomposition} shows two examples.
Expressed formally:
\[
\begin{aligned}
\RQE ::= \AQE
\mid \RQE \; \langle\mathrm{binaryOp}\rangle \; \mathrm{scalar}
\mid \mathrm{scalar} \; \langle\mathrm{binaryOp}\rangle \; \RQE \\
\mid \RQE \; \langle\mathrm{binaryOp}\rangle \; \RQE 
\mid \operatorname{aggregation}(\RQE)
\end{aligned}
\]

Inferring the optimal set of plans for a set of \AQES, instead of a set of \RQES, is safe because (a) estimating the result of an \RQE from its constituent \AQES is trivial, and (b) the \AQES constituting an \RQE model most of the computational cost of an \RQE. This is because binary operations are much cheaper than aggregations. Moreover, with a series of nested aggregations, the innermost aggregation operates on the largest volume of data and thus usually represents most of the computational cost\footnote{We realize that this is not a universal truth, but make this assumption in the interest of making \sysname's \Planner's problem tractable.}.
For simplicity, we only support \RQES where each constituent \AQE can be satisfied by \sysname.\footnote{For an \RQE where atleast one constituent \AQE is not supported, we fallback to the pre-existing \mo backend}.

\subsubsection{Generating feasible plans}
\label{subsubsec:feasible_plan_generation}

\begin{figure*}[t]
    \centering
    \includegraphics[width=0.7\linewidth]{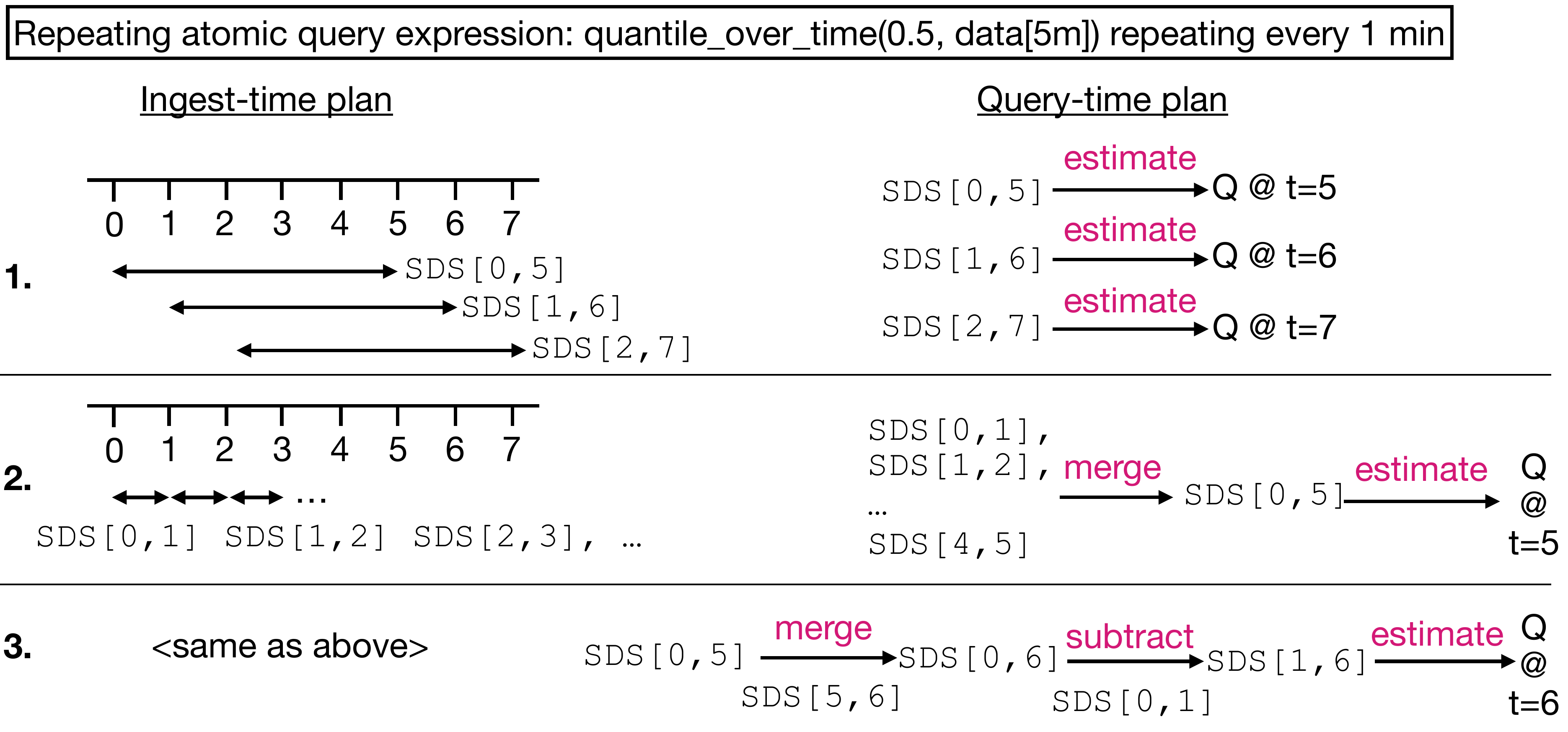}
    \caption{Illustrative plans for an example \AQE.}
    \label{fig:example_plans}
\end{figure*}

Given these simpler \AQES, we now take a few examples and explain how they can be reformulated to run on \summaries (Figure~\ref{fig:example_plans}).

\mypara{Example 1}
Consider an \AQE that repeats every 1 min: \texttt{quantile\_over\_time(0.5, cpu\_usage[5m])}.
For simplicity, assume that the \texttt{cpu\_usage} metric has only one label with only one value.
One plan to estimate this \AQE is to construct a set of \SDSS using the KLL~\cite{kll} \SummaryOp where each \SDS corresponds to an \AQE instance. Thus, we compute one KLL \SDS for \textit{each of the time ranges}: $(0, 5]$, $(1, 6]$, $(2, 7]$, ... i.e. sliding windows of size 5 min and slide 1 min. At query-time, we simply use the corresponding \SDS and estimate the median.

Another plan leverages the fact that KLL is \mergeable.
For the same \AQE, we construct \SDSS in 1 min \textit{tumbling windows} i.e. $(0, 1]$, $(1, 2]$, $(2, 3]$, and so on, at ingest-time. At query-time, we \textit{merge} 5 consecutive \SDSS to obtain a new \SDS that represents a 5 min time range. Then we use this \SDS to estimate the median. 
This plan inserts each metric value into a single \SDS and thus has lower ingest-time cost compared to the previous plan that inserts each value into 10 \SDSS. However, this plan has higher query-time cost as it must merge \SDSS before estimating a query result.

An astute reader might notice that there is a third possible plan that uses \summary \textit{subtraction}.
At ingest-time, we compute \SDSS in the same 1 min tumbling windows.
For the \textit{first} \AQE instance on $(0, 5]$, we merge 5 \SDSS, and \textit{store the merged \SDSS}. For the next \AQE instance on $(1, 6]$, we \textit{merge} the \SDSS from the previous \AQE with the \SDS on $(5, 6]$ and then \textit{subtract} the \SDS representing $(0, 1]$ from this. This results in an \SDSS representing the range $(1, 6]$, which can then estimate the query result.
While the KLL \SummaryOp does not support this operation, there may be other quantile sketches that do\footnote{It is possible that the DDSketch~\cite{ddsketch} algorithm can be modified to enable subtraction.}.

\todo{Given that we are not using HydraKLL, should we remove the whole subpopulation aware angle?}
\mypara{Example 2}
Orthogonal to the above plans that explore different time windows, let us illustrate another example. Consider the same \AQE: \texttt{quantile\_over\_time(0.5, cpu\_usage[5m])} repeating every 1 min, except the \texttt{cpu\_usage} metric has 5 labels and each label has 20 possible values, resulting in $20^5 = 3.2M$ values across all labels.
The KLL \SummaryOp can compute the median over a set of values but cannot distinguish which of the $3.2M$ subpopulations a value belongs to. We refer to this as a \SummaryOp that is non-\subpopAware.
To satisfy this query using the first plan from Figure~\ref{fig:example_plans}, we would need to maintain $3.2M$ \SDSS for every \AQE instance, wherein each \SDSS only ingests value from a specific subpopulation of the \texttt{cpu\_usage} metric.
On the other hand, there are \subpopAware{} \SummaryOp such as Hydra~\cite{hydra} and OmniSketch~\cite{omnisketch} that can ingest data from multiple subpopulations and estimate statistics on a single subpopulation i.e. a subset of the ingested data. While these specific algorithms do not support quantiles, we could imagine a \subpopAware{} \SummaryOp that can model different data-subpopulations and estimate their quantiles, such as a composite Hydra-KLL \SummaryOp. In such a case, we would only need to maintain one such \SDS for the entire \texttt{cpu\_usage} \metric.

\mypara{Generating feasible plans}
Generalizing from these examples, we define three properties for each \SummaryOp and then describe the pseudocode for generating feasible plans.

\begin{figure}[t]
    \centering
    \includegraphics[width=0.7\linewidth]{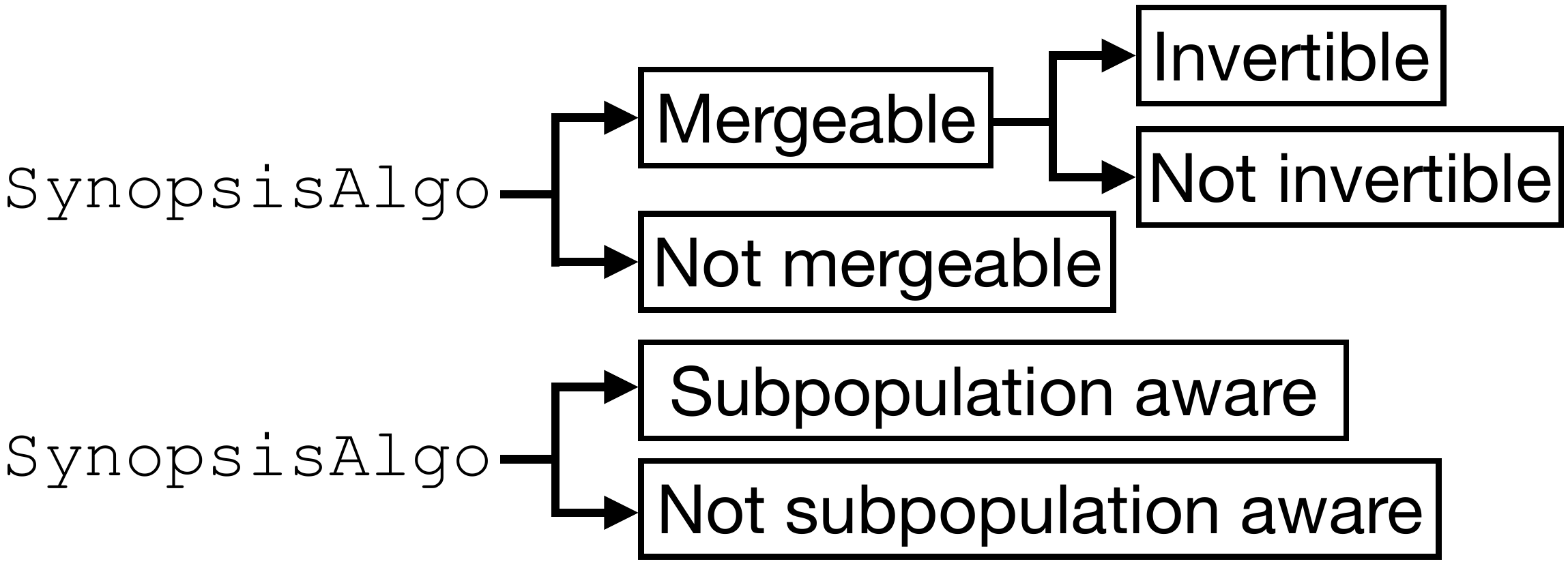}
    \caption{The \Planner taxonomizes \SummaryLib to enable generating candidate plans. 
    \zeying{I think you don't need to explain the mergability here very detail. It is the common knowledge for synopsis. subpopulation is also not a general enough case here to list, or the  counterpart as mergeability..? So taxonomy includes: a) operatability of synopsis (which is common knowledge), b) aggregation intents/statistics supported, mapping to your promql query requirements, c) synopsis working with window frameworks }}
    \label{fig:synopsis_properties}
\end{figure}

\zeying{the following paras are not logically connected}
A \SummaryOp is \subpopAware iff it can compute a statistic on different subsets of the data ingested into it. For instance, Hydra~\cite{hydra} and OmniSketch~\cite{omnisketch} are \subpopAware while HLL~\cite{hll} and KLL~\cite{kll} are not.
A \SummaryOp is \mergeable iff:
\[
\exists\, \oplus \quad \text{s.t.} \quad
SDS(D_0) \oplus SDS(D_1)
= SDS(D_0 \cup D_1).
\]
A \SummaryOp is \invertible iff:
\[
\exists\, \ominus \quad \text{s.t.} \quad
SDS(D_0) \ominus SDS(D_1)
= SDS(D_0 \setminus D_1).
\]
Figure~\ref{fig:synopsis_properties} visually depicts the relationship between these properties. An \invertible \summary is always \mergeable; and the \subpopAware property is orthogonal to the other two properties. 
A \summary may be \subpopAware but neither \mergeable nor \invertible. Another \summary may not be \subpopAware but may be both \mergeable and \invertible.

The input to the candidate plan generation algorithm is a set of \AQES and the output is a map from each \AQE to a set of feasible plans.
Each \AQE is as a tuple of (metric, spatial filter, lookback time range, statistic, group by labels, repetition interval). 

\todo{@MS polish and write as pseudocode/decision tree}
If \SummaryOp is:
\begin{packeditemize}
    \item Not \mergeable: Construct \SummaryChain with time window behavior = sliding(size=$\AQEMath.\Timevar_\lookback$, slide=$\AQEMath.\Timevar_\repeating$). Construct \EstimationDAG with only \estimateSDS operator.
    \item \Mergeable: Construct multiple \SummaryChain with time window behaviors = sliding(size=$x$, slide=$y$) where $\AQEMath.\Timevar_\lookback \bmod x == 0$, $x \bmod y == 0$, and $\AQEMath.\Timevar_\repeating \bmod y == 0$
\end{packeditemize}

If \SummaryOp is \mergeable, we can further divide it into 2 cases where \SummaryOp is \invertible or not
\begin{packeditemize}
    \item Not \invertible: Construct \EstimationDAG with transformation = merge the last $N$ windows where $N = \frac{\AQEMath.\Timevar_\lookback}{size}$.
    \item \Invertible: Construct \EstimationDAG with transformation = merge last query's result with latest \SDS and subtract oldest \SDS.
\end{packeditemize}

\begin{packeditemize}
    \item if \SummaryOp is not subpopulation-aware: construct \SummaryChain's labels to partition by = group by labels required by \AQE.
    \item if \SummaryOp is subpopulation-aware: set \SummaryChain's \SummaryOp's ``subpopulation-key'' = group by labels required by \AQE. Set  \SummaryChain's labels to partition by = empty
\end{packeditemize}

\vyas{is this candidate set "pruning" the space of possible solutions or is it just enumerating all possible things?  is there something on the optimality-tractability-efficiency tradeoff being mde here}

\subsubsection{Formulating the optimization problem}
\label{subsubsec:optimization_problem}

\mypara{Inputs}
The input is a set of \AQES $\AQESetMath = \{\AQEMath\}$ and a set of candidate plan pairs for each \AQE i.e. $\PlanSetMath_\AQEMath = \{(\summaryChainMath,\estimateDAGMath)\}$, generated from the above logic. Each candidate plan $\PlanMath \in \PlanSetMath_\AQEMath$ is a \SummaryChain and \EstimationDAG pair
that can estimate the result of \AQE $\AQEMath$.
Each \AQE has optional accuracy and latency requirements: $\accuracySLAMath_\AQEMath$, $\latencySLAMath_\AQEMath$. 

\mypara{Output}
The output of the optimization formulation is a mapping from each \AQE $\AQEMath$ to a selected $(\summaryChainMath,\estimateDAGMath) \in \PlanSetMath_\AQEMath$. Multiple \AQES may be mapped to candidate pairs containing the same
\SummaryChain, allowing the corresponding materialized \SDSS
to be shared across these \AQES.

We use two decision variables to formulate the optimization.
(1) $\decisionDeploySummaryChain_{\summaryChainMath} = 1$ if \SummaryChain $\summaryChainMath$ is deployed, and $0$ otherwise.
(2) $\decisionMapPairToAQE_{\AQEMath,\summaryChainMath,\estimateDAGMath} = 1$ if \AQE $\AQEMath$ is mapped to candidate pair $(\summaryChainMath,\estimateDAGMath)$, and $0$ otherwise.

\mypara{Internal variables}
We define four internal variables.
(1) $\ingestCostMath_{\summaryChainMath}$ denotes the ingest cost of deploying \SummaryChain $\summaryChainMath$. 
(2) $\queryCostMath_{\AQEMath,\summaryChainMath,\estimateDAGMath}$ denotes the cost of estimating the query result for \AQE $\AQEMath$ using candidate pair $(\summaryChainMath,\estimateDAGMath)$.
(3) $\accuracyFuncMath_{\AQEMath,\summaryChainMath,\estimateDAGMath}$ denotes the accuracy of estimating \AQE $\AQEMath$ using candidate pair $(\summaryChainMath,\estimateDAGMath)$.
(4) $\latencyFuncMath_{\AQEMath,\summaryChainMath,\estimateDAGMath}$ denotes the latency of estimating \AQE $\AQEMath$ using candidate pair $(\summaryChainMath,\estimateDAGMath)$.

\mypara{Objective}
The objective is to minimize the total system cost per unit time,
consisting of the ingestion cost of the materialized \SummaryChain
instances and the frequency-weighted query execution cost. For simplicity, define $\CandidateSummaryChainSetMath$ as the set of candidate \SummaryChains across all \AQES.
\[
\min_{
\decisionDeploySummaryChain,
\decisionMapPairToAQE
}
\Big(
\sum_{\summaryChainMath \in \CandidateSummaryChainSetMath}
\decisionDeploySummaryChain_{\summaryChainMath}
\times
\ingestCostMath_{\summaryChainMath}
+
\sum_{\AQEMath \in \AQESetMath}
\sum_{(\summaryChainMath,\estimateDAGMath) \in \PlanSetMath_\AQEMath}
\decisionMapPairToAQE_{\AQEMath,\summaryChainMath,\estimateDAGMath}
\times
\frequencyRQEMath_\AQEMath
\times
\queryCostMath_{\AQEMath,\summaryChainMath,\estimateDAGMath}
\Big)
\]
The formulation accounts for sharing when multiple selected candidate pairs use the same \SummaryChain as the ingest cost of that \SummaryChain is incurred only once. Different \SummaryChains may themselves have common operators. In the interest of simplicity and tractability, we do not model such partial sharing between distinct \SummaryChains, and therefore may overestimate their combined ingest cost.
Here $\frequencyRQEMath_\AQEMath = \frac{1}{\Timevar_\AQEMath^\repeating}$

\mypara{Constraints}
Every \AQE must be mapped to exactly one candidate $(\SummaryChain,\EstimationDAG)$ pair (Eq.~\eqref{eq:cover}). An \AQE can be mapped to a candidate pair only if its \SummaryChain is deployed (Eq.~\eqref{eq:deploy}). The selected candidate pair must also satisfy the \AQE's accuracy~\eqref{eq:accuracy} and latency~\eqref{eq:latency} requirements.

\begin{align}
    \sum_{(\summaryChainMath,\estimateDAGMath) \in \PlanSetMath_\AQEMath}
    \decisionMapPairToAQE_{\AQEMath,\summaryChainMath,\estimateDAGMath}
    &= 1
    && \forall \AQEMath \in \AQESetMath
    \label{eq:cover}\\
    \decisionMapPairToAQE_{\AQEMath,\summaryChainMath,\estimateDAGMath}
    &\le
    \decisionDeploySummaryChain_{\summaryChainMath}
    && \forall \AQEMath \in \AQESetMath,\ 
    (\summaryChainMath,\estimateDAGMath) \in \PlanSetMath_\AQEMath
    \label{eq:deploy}\\
    \decisionMapPairToAQE_{\AQEMath,\summaryChainMath,\estimateDAGMath} = 1
    &\implies
    \accuracyFuncMath_{\AQEMath,\summaryChainMath,\estimateDAGMath}
    \ge \accuracySLAMath_{\AQEMath}
    && \forall \AQEMath \in \AQESetMath,\ 
    (\summaryChainMath,\estimateDAGMath) \in \PlanSetMath_\AQEMath
    \label{eq:accuracy}\\
    \decisionMapPairToAQE_{\AQEMath,\summaryChainMath,\estimateDAGMath} = 1
    &\implies
    \latencyFuncMath_{\AQEMath,\summaryChainMath,\estimateDAGMath}
    \le \latencySLAMath_{\AQEMath}
    && \forall \AQEMath \in \AQESetMath,\ 
    (\summaryChainMath,\estimateDAGMath) \in \PlanSetMath_\AQEMath
    \label{eq:latency}
\end{align}

\mypara{Cost and accuracy modeling}
To estimate the ingest cost, query cost, accuracy, and latency (required by the above optimization), we combine empirical per-operation measurements from an offline benchmark with an analytical cost model. 
We measure the CPU time for four operations: inserting an item into an \SDS, merging two \SDSS, subtracting one \SDS from another, and estimating a statistic from an \SDS. Then we scale these up based on the analytical cost model to provide the required measurements for the optimization problem.

\todo{add cost model 2 and 3}
The ingestion cost of a \SummaryChain depends on the number of items inserted into an \SDS and the number of parallel \SDSS being maintained. The former is estimated using the cardinality of the stream and how often an item is inserted\footnote{\Mo deployments usually ``pull'' data at regular time intervals.}, while the latter can be calculated based on the time windows. A sliding window of size $x$ and slide $y$ implies $x/y$ parallel windows being maintained at any time. The query cost of a \EstimationDAG depends on the number of merge, subtract, and estimation options. For e.g. the three plans from the example in \S\ref{subsubsec:feasible_plan_generation}, would cost (a) 1 estimation; (b) $\frac{\AQE.\Timevar_\lookback}{x}$ merges $+$ 1 estimation; and (c) 1 merge $+$ 1 subtract $+$ 1 estimation, respectively. $\AQE.\Timevar_\lookback$ is the time range queried by the \AQE. The memory at ingest and query time can be modeled analytically too.

To model accuracy, we construct \SDSS with different \SummaryOp{s}, configurations and run them on diverse data distributions (e.g. Zipf with different skewness) and obtain empirical accuracy measurements. We also measure how accuracy changes when two \SDSS are are merged, and use this to estimate the accuracy of a candidate plan $(\summaryChainMath, \estimateDAGMath)$.

\subsubsection{Putting everything together}

As shown in Figure~\ref{fig:planner_design}, the \Planner first decomposes each \RQE into its constituent \AQES. Then, it generates candidate plans, where each plan is a pair of a \SummaryChain and \EstimationDAG instance. Finally, it formulates and solves an optimization problem using empirical cost and accuracy estimates. The \Planner outputs the optimal set of plans to deploy, and a map from each \AQE to a deployed plan.

\subsection{\sysname's data plane}

\todo{up-level \summary-first materialized views across the repeating query workload}

\sysname's data plane is responsible for computing \SDSS from raw metrics data ($\summarizeFuncMath$), storing them, and then using them to estimate query results ($\estimateFuncMath$). These are the responsibilities of the \Summarizer, \StorageEngine, and \QueryEngine. 

\mypara{\Summarizer}
The \Summarizer takes a stream of raw \metrics as input and produces a stream of \SDSS. It must be able to handle high throughput of \metrics, have low freshness delay, and must be extensible to compute new types of summaries as the \SummaryLib expands. We design the \Summarizer to continuously process \metrics as they arrive, and incrementally compute \SDSS. This naturally enables ingest-time precomputation, allowing the queries to use \SDSS that have already been materialized at ingest-time; and naturally applies to \summaries since they can be computed in a streaming manner.

We leverage a pre-existing abstraction from existing stream processors~\cite{flink, arroyo}: the dataflow DAG, that makes it easy to extend to new \SummaryOp{s}. The \Summarizer translates the summarization function $\summarizeFuncMath$ into a DAG with filtering, grouping, windowing, and \SDS construction operators. 
If several \summaries require the same filter or grouping operation, the \Summarizer executes that operation once and fans its output out to downstream operators, avoiding redundant work.

\mypara{\StorageEngine}
The \StorageEngine persists computed \SDSS and organizes \SDSS according to their \SDSscope and \SDSlogic. The scope identifies the metric, time range, filters, and partitioning labels relevant to a query, while the logic determines whether summaries can be safely combined and how their payloads should be interpreted. 
Treating the \SDSpayload as opaque bytes makes the design extensible: supporting a new sketch or summary operator requires defining its scope, logic, and payload representation, rather than introducing a new storage subsystem.
The \StorageEngine leverages the fact that \mo queries mostly operate on recent data. It maintains a bounded in-memory circular buffer containing recent \SDSS for fast access, while asynchronously writing older, completed time partitions to durable storage. A query first accesses summaries in memory and retrieves persisted summaries only when its time range requires them. This design keeps the common recent-data path fast, bounds memory use, and still supports historical queries.

\mypara{\QueryEngine}
The \QueryEngine intercepts queries sent by an application and answers queries using the materialized \SDSS. By avoiding scans over raw \metrics, it substantially reduces both query latency and query-time compute cost.
The \QueryEngine has three requirements. It must provide substantial latency and compute savings over querying the underlying \mo system, it must be extensible to new queries  and \SummaryOp{s}, and it must gracefully fall-back for unsupported queries.
On receiving a query, the \QueryEngine parses it and looks up the corresponding \EstimationDAG instance. If no \EstimationDAG is found, it immediately forwards them to the existing \mo backend. If found, it uses the \EstimationDAG to infer the \SDSS to read, the merge/subtraction operators to apply and what to estimate from the \SDSS.
If an \EstimationDAG requires subtraction, it keeps track of the previous query instance's result. 
The \QueryEngine can provide significant cost-latency benefits because it operates over compact \SDSS not raw data. Moreover, most \mo queries look at recent data that is already in memory in the \StorageEngine. The \QueryEngine is easily extensible to new \summaries as the core logic is agnostic to the implementation of a particular \summary. In fact, the \Summarizer and \QueryEngine share a common \summary library implementation.

\section{Implementation}
\label{sec:implementation}

Next, we describe the \sysname implementation. \sysname   and experimental harness  is open source at \href{https://github.com/ProjectASAP/ASAPQuery}{https://github.com/ProjectASAP/ASAPQuery}.

\mypara{\Planner} The \Planner is implemented in 3k LOC of Python and uses the \texttt{py-promql-parser}~\cite{promql-parser-py} library to parse \PromQL queries into ASTs, define query templates, and match these against ASTs.The \Planner outputs the \PrecomputeConfig and \InferenceConfig as YAML files. 

\mypara{\Summarizer} The \Summarizer is implemented using  Arroyo~\cite{arroyo} a streaming engine, with  4k LOC of Python and Rust. The \Summarizer converts the \PrecomputeConfig to streaming pipelines in Arroyo's SQL dialect. We implement \SketchLib and individual \SketchOp as a Rust User Defined Function (UDF) to be used by Arroyo. The \Summarizer uses Python scripts to call Arroyo's API to configure pipelines. To enable Arroyo to interoperate with \Prom, we leverage \Prom' {\tt remote\_write} capability~\cite{remote-write}. %
We implement a custom source in Arroyo that understands the \Prom {\tt remote\_write} protocol. The \Summarizer sends \SDSS  to the \QueryEngine over Kafka~\cite{kafka}.

\mypara{\QueryEngine} The \QueryEngine is implemented in 15k LOC of Rust. The core  library implements query estimation logic for various types of sketches. We define a standard set of Rust traits in the \QueryEngine, making it easy to implement and plug in a new sketch to the core query estimation logic. Currently, \sysname implements the following \SketchLib: 
(1) KLL~\cite{kll}, (2) CountMinSketch~\cite{cmsketch}, (3) Hydra~\cite{hydra} sketch on top of KLL~\cite{kll}, (4) Exact accumulator for \texttt{rate} \& \texttt{increase}, (5) Exact accumulator for \texttt{min} \& \texttt{max}, and (6) Exact accumulator for \texttt{sum} \& \texttt{count}.
We use the \texttt{rdkafka}~\cite{rdkafka} library to ingest \SDSS from the \Summarizer, and the \texttt{promql-parser}~\cite{promql-parser-rs} library to parse queries.
The \QueryEngine implements a subset of \Prom' HTTP API~\cite{prom-http-api} to receive queries from \Grafana or any other \mo application, and uses the same HTTP API to send unsupported queries to \Prom.

\section{Evaluation}
\label{sec:results}

We evaluate \sysname on a combination of real and synthetic workloads. First, we describe our evaluation setup. Then, we conduct an end-to-end  evaluation at a modest scale to show the functional value of integrating \sysname in a \mo deployment. 
 Then, we measure cost-latency-scale profiles of different \RQE types to inform  ``what-if'' analysis experiments for diverse deployment scenarios.

\subsection{Setup}

\mypara{Baselines}
We deploy \sysname and verify its backwards-compatibility with two  production \mo systems: \Prom~\cite{prometheus} and \VM~\cite{victoria-metrics}. 
 In the interest of brevity, we  only  show the results for  \Prom. 
We use these systems out-of-the-box and do not perform any tuning.

\mypara{Measures of success} We compare \sysname vs. baseline on querying cost and latency. We assume a simple self-hosted cloud billing model where cost depends on the number of CPU cores allocated to a machine. Thus, we measure query cost using the p99 CPU usage of \sysname vs. baselines, ignoring \metrics ingestion cost.

\mypara{Data and Query workload} We use a combination of real metrics  data from  prior work~\cite{alibaba-traces-2021, google-traces-2011} and a synthetic data generator. 
From the,  Alibaba microservices 2021 and 2022 dataset~\cite{alibaba-traces-2021, alibaba-traces-2022}, we use the ``node'' and ``MSResource'' tables that represent \metrics data.  
From the  Google Borg cluster data~\cite{google-traces-2011}, we use the ``task\_usage'' measurements. We also implement a synthetic  data generator that allows us to control the number of metadata labels and the cardinality of each label.  
We ingest \metrics data every 1 second.

\mypara{Testbed} We run our experiments on the ``m510'' node in the Cloudlab Utah cluster~\cite{cloudlab}. Each node has 8 Intel Xeon D-1548 (2.0 GHz) cores and 64GB of RAM. We run each experiment using 2 nodes to for accurate measurement -- one runs \sysname or the baseline, while the other runs the \metrics data exporters.

\subsection{End-to-end validation}

\begin{figure}[t]
    \centering
    \includegraphics[width=0.8\linewidth]{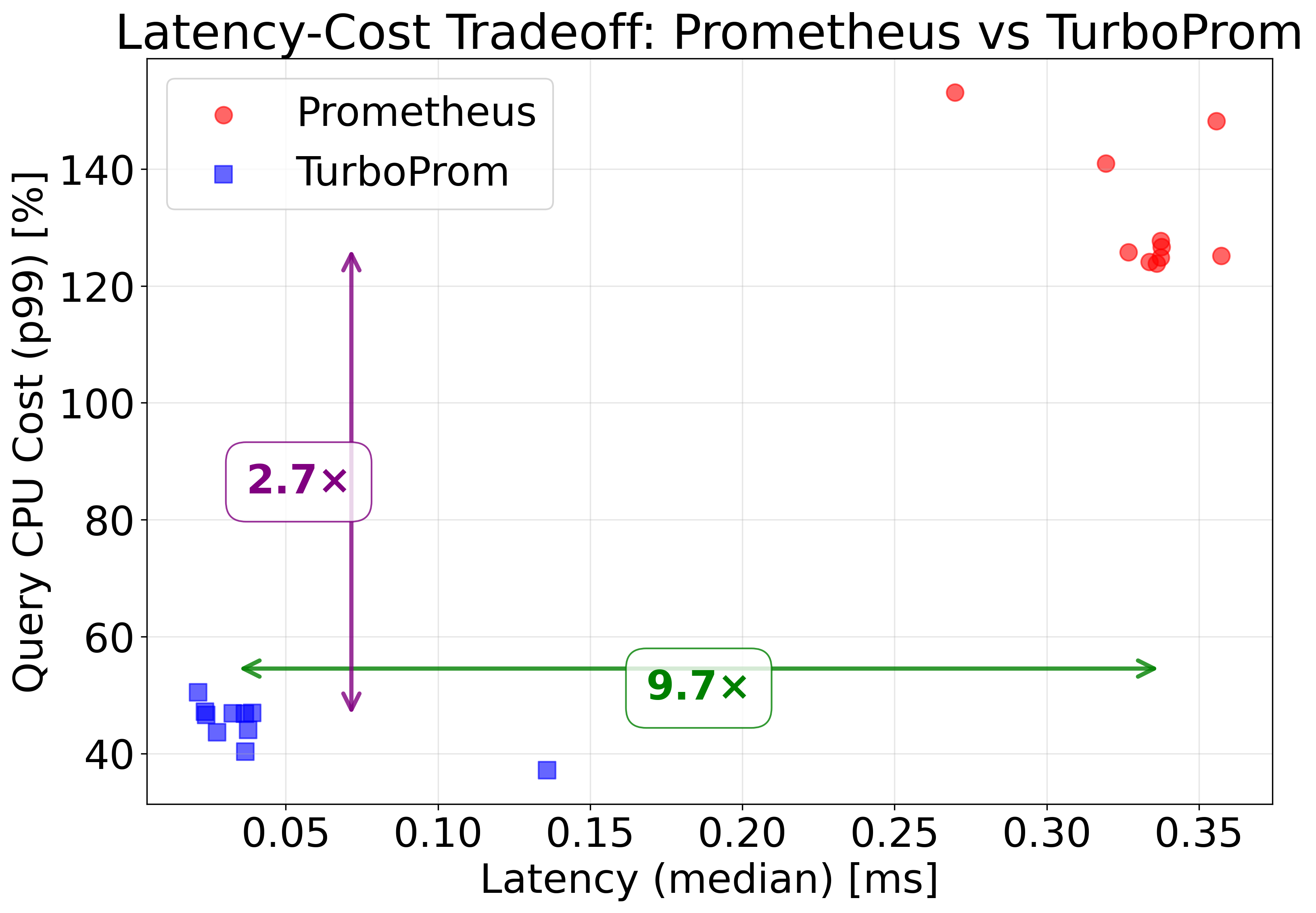}
    \vspace{-0.3cm}
    \caption{Across diverse randomly generated workloads, \sysname reduces median query latency and p99 querying cost, compared to state-of-the-art \Prom.}
      \vspace{-0.3cm}
    \label{fig:result_punchline}
\end{figure}

Recall that any \PromQL \queryWorkload consists of multiple \RQES, where each \RQE consists of one or more $\AtomicQuery$ units, each having one or more aggregations. To evaluate \sysname vs \Prom, we construct a set of 10 \textit{randomly generated} workloads using the $\AtomicQuery$ building blocks discussed earlier (\S~\ref{sec:background}), consisting of \SimpleTemporal, \SimpleSpatial, and \OneSpatialTemporal. Since \SimpleSpatial queries run on a single time instant, we set these to repeat every 1 sec. \SimpleTemporal and \OneSpatialTemporal aggregate data across time. Inspired by workload patterns from real \Prom deployments, we fix these to aggregate 15 min of data, and repeat every 1 min. We start with a modest deployment generating 8k measurements/sec and run each query workload for 20 min\footnote{We set sketch parameters to reasonable defaults; $k=20$ for KLL and $3\times 512$ for CountMinSketch}. For each \queryWorkload, we measure the median latency and p99 querying cost. Figure~\ref{fig:result_punchline} shows that across a \textit{diverse range of random workloads}, \sysname simultaneously provides both cost and latency benefits over a state-of-the-art \mo system. 
 We observe similar benefits compared to \VM too (not shown).

\begin{figure}[t]
    \centering
    \includegraphics[width=0.7\linewidth]{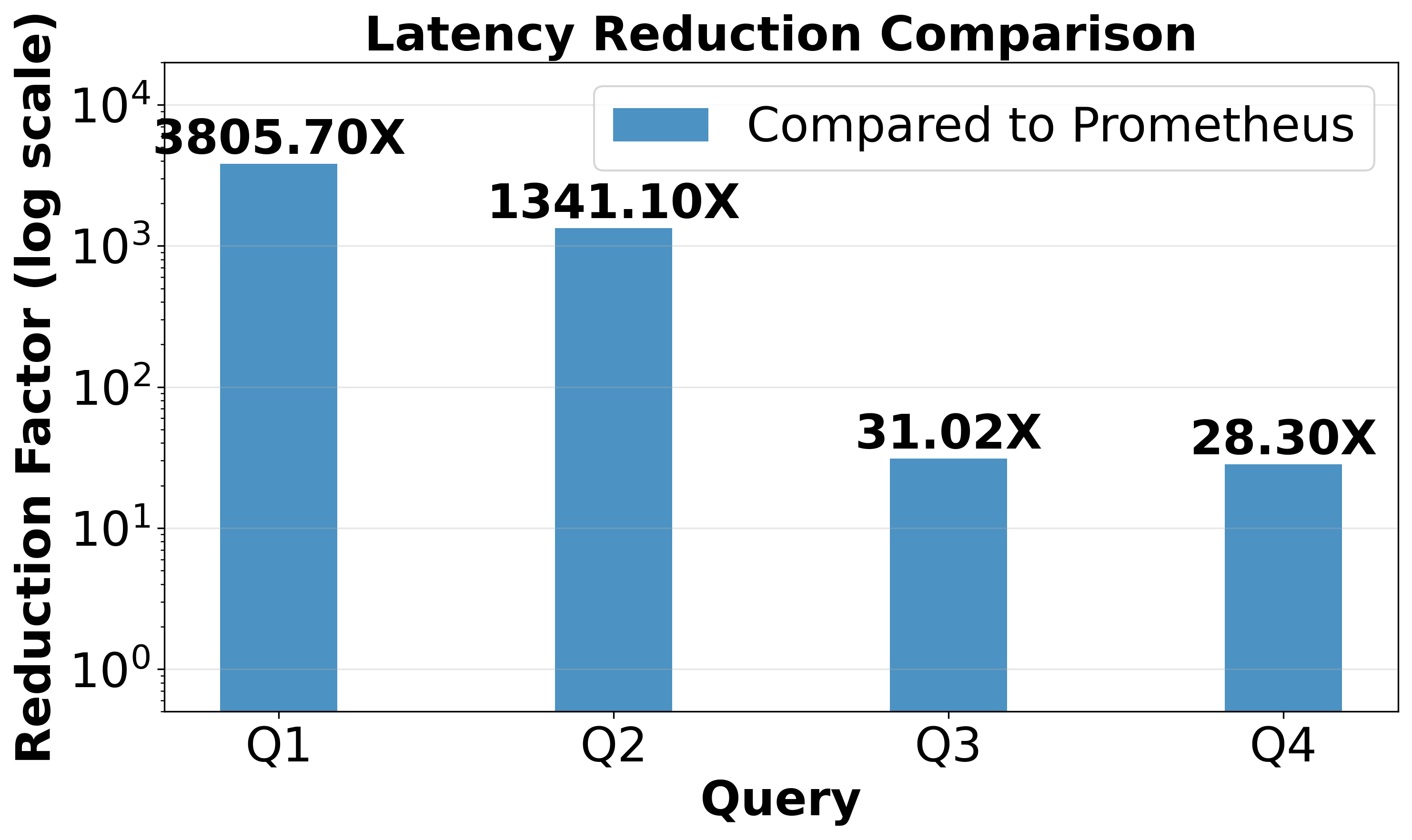}
      \vspace{-0.3cm}
    \caption{\sysname provides benefits across representative $\AtomicQuery$ instances}
      \vspace{-0.5cm}\label{fig:result_per_atomic_query}
\end{figure}

\begin{figure}[t]
    \centering
    \includegraphics[width=0.9\linewidth]{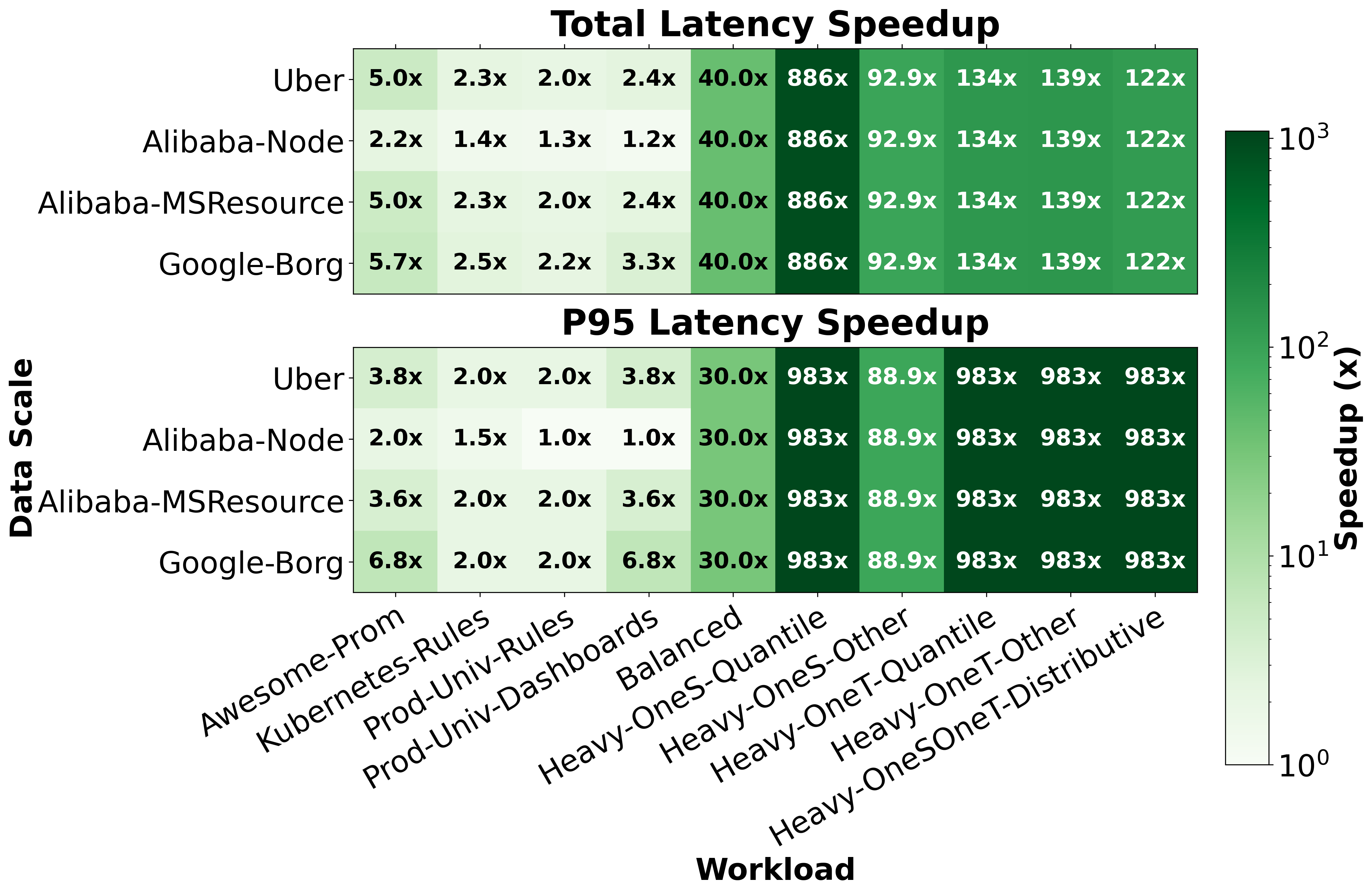}
      \vspace{-0.4cm}
    \caption{Estimating  \sysname's benefits for total and p95 latency on diverse scenarios.}
    \label{fig:result_heatmaps}
     \vspace{-0.5cm}
\end{figure}

\subsection{What-if Scenarios}
 Given this functional validation, we would like to estimate \sysname's benefits for hyperscalar enterprise deployments. An ideal evaluation requires the \queryWorkload of a hyper-scalar and their data \textit{shape} and \textit{scale}. Unfortunately, there is no public information on enterprise \queryWorkloads, and running these workloads at large scale is prohibitively expensive.  %
 To this end, we build a scenario generator based on profiling diverse query patterns and scaling behaviors. Then, we extrapolate the potential  benefits for different offered loads.

\mypara{Profiling  $\AtomicQuery$  patterns} 
First, we profile  \sysname's latency benefits vs baseline \textit{individually} for different $\AtomicQuery$, at a much larger scale to stress out the query path.
We construct the \queryWorkload such that each individual query processes approximately 1M data points and run 10 query repetitions, without any potential reuse across time. We run four different $\AtomicQuery$: \SimpleSpatial \texttt{quantile}, \SimpleSpatial \texttt{sum}, \SimpleTemporal \texttt{quantile}, and a \OneSpatialTemporal with distributive \texttt{sum} on \texttt{sum\_over\_time} aggregation. 

Figure~\ref{fig:result_per_atomic_query} shows the latency reduction of \sysname vs. \Prom. While each $\AtomicQuery$ has 1M input data points, the semantics vary per query, resulting in different benefits. 
A \SimpleSpatial query provides the most benefits as it aggregates 1M data points (at a single time instant) and produces a query result with 40 keys. Thus, each key aggregates 1M/40 = 25K data points. A \SimpleTemporal query produces a query result with 1000 keys where each key aggregates only 1000 points, 25X lesser than the \SimpleSpatial case.

\begin{figure}[t]
    \centering
    \begin{subfigure}{\linewidth}
        \centering
        \includegraphics[width=0.9\linewidth]{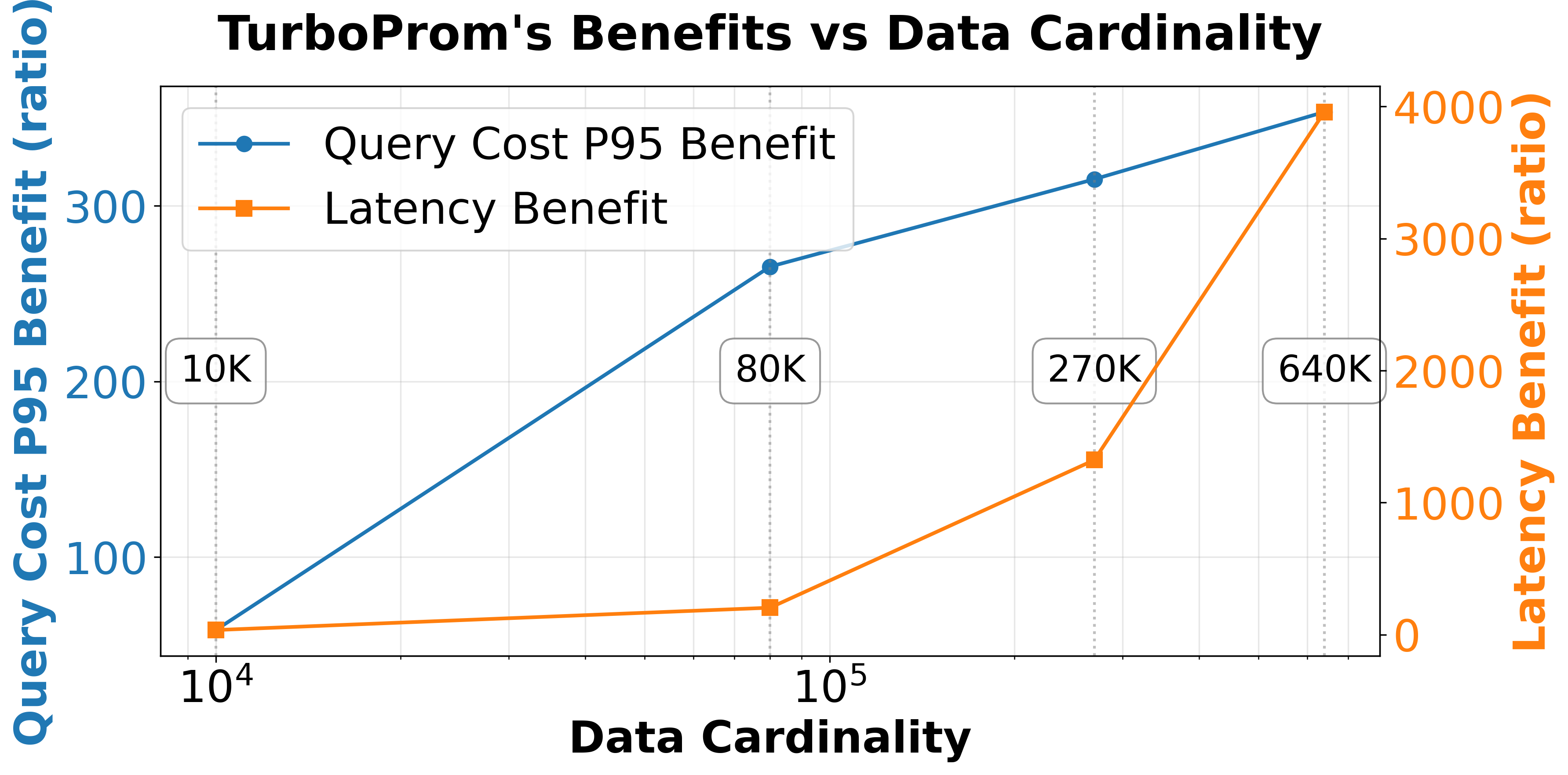}
        \caption{Varying  cardinality for \SimpleSpatial \texttt{quantile} $\AtomicQuery$}
    \end{subfigure}
    \begin{subfigure}{\linewidth}
        \centering
        \includegraphics[width=0.9\linewidth]{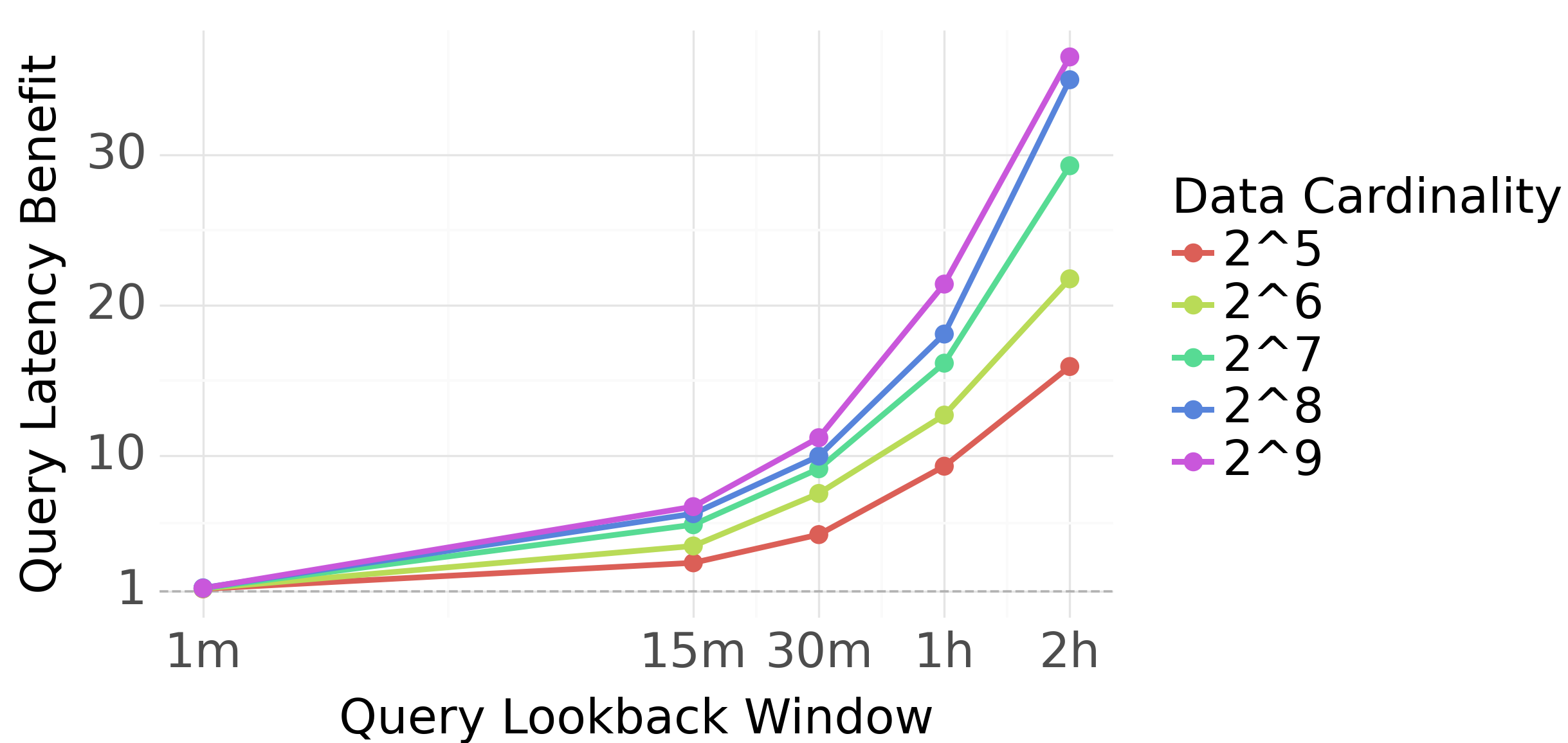}
        \caption{Varying input data cardinality and query lookback window for \SimpleTemporal \texttt{quantile} $\AtomicQuery$}
    \end{subfigure}
     \vspace{-0.5cm}
    \caption{\sysname's latency and cost benefits over \Prom increase with increasing query load}
    \label{fig:results_vs_query_load}
     \vspace{-0.5cm}
\end{figure}

\mypara{Improvements with scale}
We evaluate \sysname's benefits with increasing query load on two candidate perceived-to-be-expensive $\AtomicQuery$ - a \SimpleSpatial \texttt{quantile} query, and a \SimpleTemporal \texttt{quantile\_over\_time} query. Figure~\ref{fig:results_vs_query_load} shows the result for these.
Query load for a \SimpleSpatial query is determined only by the data cardinality as it aggregates spatially, not temporally. Thus, we increase the query load by increasing data cardinality and evaluate query cost reduction and latency reduction. To scale up cardinality, we run 10 data generators, each producing \metrics with 3 labels, with label cardinality varying in $C=\{10, 20, 30, 40\}$. The resulting cardinality is $10\times C^3$ = $\{10K, 80K, 270K, 640K\}$.
Query load for a \SimpleTemporal query depends on the query lookback window i.e. the time range queried, and the data cardinality. We evaluate on lookback windows $\{1m, 15m, 30m, 60m, 120m\}$ (data is ingested once per second) and cardinality in the range $\{2^5, ..., 2^9\}$. \sysname's latency benefit over baseline \Prom increases with both query lookback window and data cardinality.

\mypara{Extrapolating to   deployments} 
 We design 10 \textit{workload mixes}, each characterized by the supported \RQES in a \queryWorkload and a distribution of $\AtomicQuery$ units in the supported \RQES. We design 4 of these using the datasets \datasetAwesomeProm, \datasetKubernetes, and \datasetCMU. We design 5 more by creating \textit{workload mixes} that contain 80\% of a specific type of an $\AtomicQuery$ (such as $\SimpleSpatial$ with \texttt{quantile} or \SimpleTemporal with \texttt{rate}); with other $\AtomicQuery$ types having 5\% each. The last workload mix is evenly balanced across all $\AtomicQuery$.
 For the data shape (i.e.,  the data cardinality and number of data points aggregated by each query), we rely on  public blogs and \metrics data released by 3 hyperscalars: Uber~\cite{uber-blog-1}, Alibaba~\cite{alibaba-traces-2021, alibaba-traces-2022} and Google~\cite{google-traces-2011}.
  We create four realistic \textit{data shapes}, defined by number of data points $N$ for \SimpleSpatial and cardinality $C$ for \SimpleTemporal (for this, we set $T=30$ data points as there is no public info). For $N, C$, we set (a) Uber $(N=3M, C=100K$) (b) Alibaba $(N=390K, C=13k)$; $(N=2.8M, C=96k)$, (c) Google $(N=5.4M, C=180K$).
  
 Finally, for  each combination of a (\textit{workload mix}, \textit{data shape}), we estimate the original query latencies and \sysname's benefits in reducing it, by linearly extrapolating from our experiments (that were run at smaller scale).

Figure~\ref{fig:result_heatmaps} shows two heatmaps estimating the reduction in the \textit{total query latency}, and the \textit{p95 query latency}, for 10x4 combinations of workload mixes and data shapes. We see that \sysname provides consistent benefits across a diverse range of workload mixes, both real and synthetic for hyper-scale deployments. For instance, for overall latency speedup, we see benefits  ranging from 1.2X in the worst case to 900X depending on the expected workload.

\section{Other Related Work}
We discussed closest related work in \S\ref{sec:background}. Here we discuss other related work in observability.

\myparatight{Proprietary industry solutions}
  In light of the cost-scale-latency tradeoffs of \mo, many hyperscalars have developed proprietary architectures and data store technologies~\cite{uber-blog-1,netflix-datamesh,monarch,datadog-husky,gorilla}. To the best of our knowledge they are still based on a classic data and query model. We believe these solutions can also benefit from the ideas in \sysname. 

  \myparatight{Horizontal scaling for \mo} 
   In this work, we focused on scaling query processing on a single node. There are complementary approaches for horizontal scalability~\cite{thanos}. An interesting future work is exploring the applicability of key ideas in \sysname to scale-out deployments, including for distributed querying and long-term storage.

\myparatight{Query optimization} There is a large body of work in the database and data streaming communities~\cite{catalog-stream} for query planning and optimization. While there are some point efforts to include some sketches (e.g., HLL for count distinct) in some databases, there has not been a concerted effort on query planning in the context of summary-based computation.

\myparatight{Log and Trace observability} 
While we focused on \mo in this paper, there are other aspects of observability  like  logs and tracing that  also  face significant cost and scaling bottlenecks. There are efforts for more compact logging~\cite{clp} and reducing the cost of tracing~\cite{hindsight}. One direction of future work is to understand if, and how, our insights generalize to log and trace workloads.

\myparatight{Approximate query processing (AQP)}
AQP in databases is not a new idea~\cite{pass, aqp++, verdictdb, blinkdb}. However, these do not directly focus on \mo workloads or deployment constraints that we address in \sysname.

\section{Discussion}
\label{sec:discussion}
\myparatight{Opportunities to improve \sysname} 
One idea is to expand \SketchLib with more capabilities and improve our heuristic algorithm. We can also consider optimizing the \SketchOp parameters. These may benefit from recent advances in AI-driven research systems~\cite{berkeleyadrs,openevolve}. 
 Finally, we plan to explore the benefits of \sysname in a horizontally-scaled or cluster deployment~\cite{thanos}.

\myparatight{Other deployment models} 
 This work focuses on {\em self hosted} Prometheus. Hosted  Prometheus/SaaS solutions (e.g., Datadog) often charge based   on {\em  data ingest}. While customers cannot directly benefit from \sysname, the vendor  can use our  ideas and transfer cost benefits to customers. 

 \myparatight{Beyond \mo}  Our work was motivated by an imminent need for reducing the cost-latency concerns in \mo with de-facto tools like Prometheus. However, the key concepts here can apply to more generally to other timeseries databases (e.g.,~\cite{timescale,influx,clickhouse} ), streaming systems (e.g.,~\cite{flink, arroyo}), and other observability stacks (e.g.,~\cite{elastic,signoz}).

\section{Conclusions}

As systems scale in size and complexity,  monitoring the health of the systems has itself become a significant bottleneck and cost sink~\cite{high-obs-cost-industry-2}.   Despite years of   work, the cost-latency tradeoffs for \mo  are still a challenge. By taking a first principles approach to understand the nature of the data and queries, we identify a new opportunity to recast the problem using \sysnameView. We show  this idea can enable  a practical system  ``drop in'' accelerator  that   yields significant savings. As we enter  an era for large-scale systems where ``throwing more machines'' is  no longer sufficient,  the key ideas in \sysname---the  \sysnameView   and the named-data summary  abstraction--- are likely more broadly applicable to other observability and data processing systems at scale.

\bibliographystyle{plain}
\bibliography{refs/online, refs/sketches, refs/prior_work, refs/misc}

\end{document}